\documentclass[journal,twoside,web]{ieeecolor}
\usepackage{generic}
\usepackage{cite}
\usepackage{amsmath,amssymb,amsfonts}
\usepackage{algorithmic}
\usepackage{graphicx}
\usepackage{algorithm,algorithmic}
\usepackage{hyperref}
\usepackage{bbding}
\usepackage[table]{xcolor}
\usepackage{diagbox}
\usepackage{graphicx}
\usepackage{amsmath}
\usepackage{amssymb}
\usepackage{booktabs}
\usepackage{framed,multirow}
\usepackage{multirow}
\usepackage{diagbox}
\usepackage{url}
\usepackage{diagbox}
\usepackage{makecell}
\usepackage{enumerate}
\usepackage{array}
\usepackage{graphicx}
\hypersetup{hidelinks=true}
\usepackage{textcomp}
\def\BibTeX{{\rm B\kern-.05em{\sc i\kern-.025em b}\kern-.08em
    T\kern-.1667em\lower.7ex\hbox{E}\kern-.125emX}}
\begin{document}
\title{Deformable Medical Image Registration with Effective Anatomical Structure Representation and Divide-and-Conquer Network}
\author{Xinke Ma, Yongsheng Pan, Qingjie Zeng, Mengkang Lu, Bolysbek Murat Yerzhanuly, Bazargul Matkerim, and Yong Xia
\thanks{
This work was supported by the National Natural Science Foundation of China (62171377), the Key R\&D Program of Shaanxi Province (2022GY-084), the Fundamental Research Funds for Central Universities (5000230376), and the Ningbo Clinical Research Center for Medical Imaging (2021L003, Open Project 2022LYKFZD06).
%
%
({\em X. Ma and Y. Pan contributed equally. Corresponding author: Y. Xia})}
\thanks{X. Ma, Y. Pan, Q. Zeng, M. Lu, and Y. Xia are with the National Engineering Laboratory for Integrated Aero-Space-Ground-Ocean Big Data Application Technology, School of Computer Science and Engineering, Northwestern Polytechnical University, Xi'an 710072, China. (e-mail: \{maxxk, yspan, qjzeng, lmk\}@mail.nwpu.edu.cn; yxia@nwpu.edu.cn)
}
\thanks{B. Yerzhanuly is at Northwestern Polytechnical University's Kazakhstan branch. (e-mail: muratbolysbek@mail.nwpu.edu.cn)}
\thanks{B. Matkerim is with the Al-Farabi Kazakh National University, Almaty 050040, Kazakhstan. (e-mail: Bazargulmm@gmail.com)}
%
%
%
}
\maketitle

\begin{abstract}
Effective representation of Regions of Interest (ROI) and independent alignment of these ROIs can significantly enhance the performance of deformable medical image registration (DMIR). However, current learning-based DMIR methods have limitations. Unsupervised techniques disregard ROI representation and proceed directly with aligning pairs of images, while weakly-supervised methods heavily depend on label constraints to facilitate registration. To address these issues, we introduce a novel ROI-based registration approach named EASR-DCN. Our method represents medical images through effective ROIs and achieves independent alignment of these ROIs without requiring labels. Specifically, we first used a Gaussian mixture model for intensity analysis to represent images using multiple effective ROIs with distinct intensities. Furthermore, we propose a novel Divide-and-Conquer Network (DCN) to process these ROIs through separate channels to learn feature alignments for each ROI. The resultant correspondences are seamlessly integrated to generate a comprehensive displacement vector field. Extensive experiments were performed on three MRI and one CT datasets to showcase the superior accuracy and deformation reduction efficacy of our EASR-DCN. Compared to VoxelMorph, our EASR-DCN achieved improvements of 10.31\% in the Dice score for brain MRI, 13.01\% for cardiac MRI, and 5.75\% for hippocampus MRI, highlighting its promising potential for clinical applications. {\color{black}The code for this work will be released upon acceptance of the paper.}
\end{abstract}

%
\begin{figure}
\centering
\includegraphics[width=0.49\textwidth]{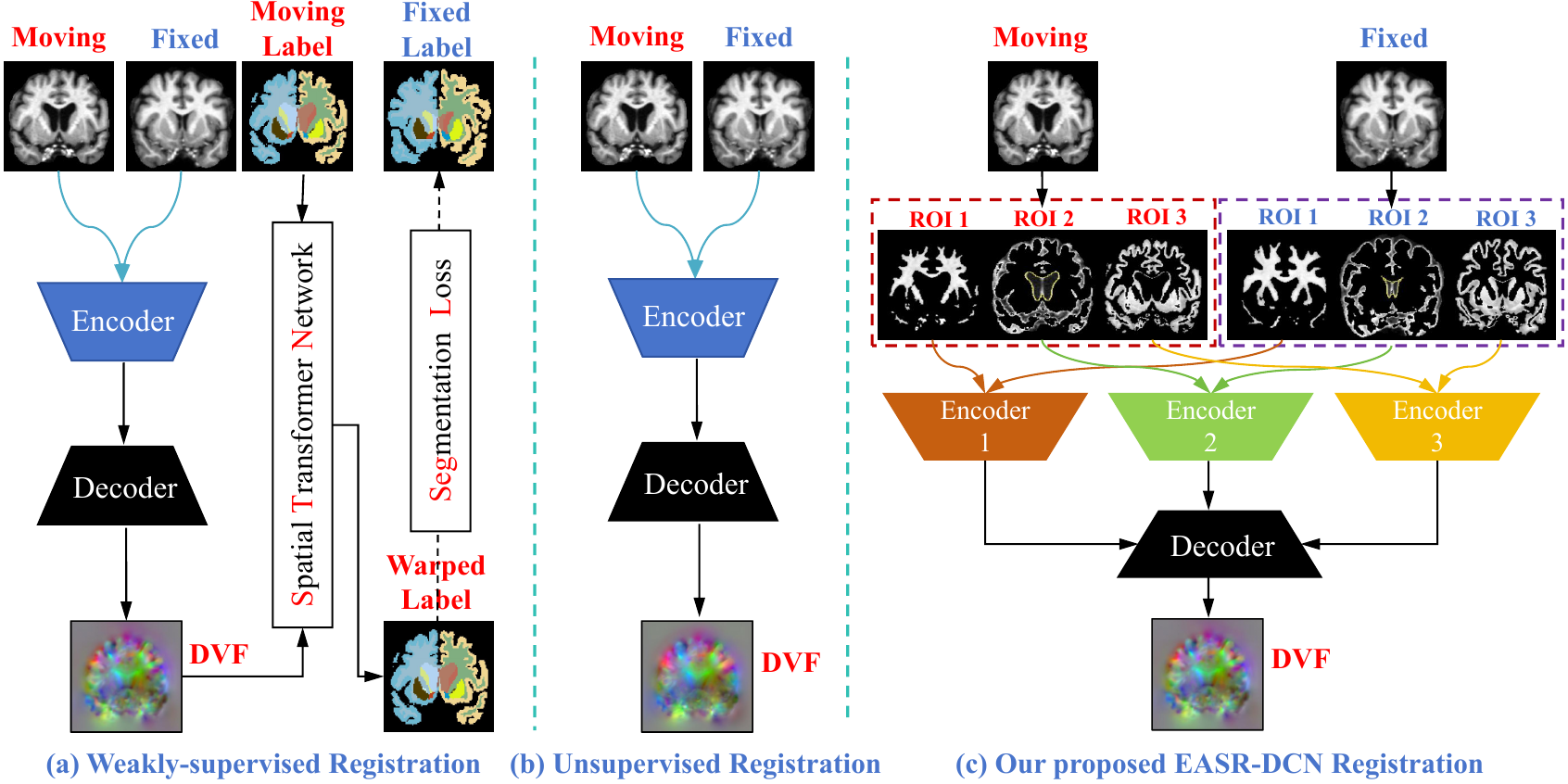}
\caption{Comparison of unsupervised and weakly-supervised registration methods with our EASR-DCN model: (a) The weakly-supervised registration method uses the whole image in a \textbf{S}ingle-\textbf{c}hannel \textbf{N}etwork (ScN) but with labels for DVF generation; (b) The unsupervised registration method uses the whole image in ScU for DVF generation without labels; (c) Our EASR-DCN divides the image into multiple effective ROIs with distinct intensities and feeds them into the DCN to independently learn the feature alignment of ROIs for DVF generation. 
This effectively avoids interference between the ROIs.
}
\label{img1}
\end{figure}
\section{Introduction}
\label{sec:intro}
\IEEEPARstart{D}{eformable} Medical Image Registration (DMIR) is crucial in medical image analysis.
It involves applications such as medical image segmentation \cite{10042453}, tracking temporal anatomical changes \cite{zheng2021symreg}, and analyzing anatomical variations across populations \cite{yang2019shared}. The primary goal of DMIR is to establish a dense, nonlinear correspondence between two images to estimate their alignment transformation \cite{zheng2024residual,10621700}.

Traditional DMIR techniques \cite{avants2008symmetric,ma2020pointpr} avoid paired training, but rely on slow iterative optimization, limiting real-time use. Deep learning advances, using Convolutional Neural Networks (CNNs) or Transformers, have accelerated and improved DMIR accuracy. Learning-based DMIR methods fall into supervised and unsupervised categories.

\textbf{Supervised DMIR methods} are divided into fully and weakly supervised approaches. Fully supervised methods depend on ground-truth \textbf{D}isplacement \textbf{V}ector \textbf{F}ields (DVFs) for registration. However, manually acquiring these DVFs is often prohibitively expensive, so they are typically generated synthetically or derived from conventional registration methods. Due to these limitations in fully supervised approaches, researchers are increasingly focusing on weakly supervised and unsupervised methods.
Weakly supervised learning-based DMIR methods, as illustrated in Fig. \ref{img1} (a), first automatically or manually segment anatomical structures and then use these segmentation labels to guide the registration network in aligning corresponding anatomical regions between moving and fixed images. Among them, methods such as WS-VM \cite{balakrishnan2019voxelmorph}, RsegNet \cite{qiu2021rsegnet}, SUITS \cite{blendowski2021weakly}, WS-GM \cite{tan2024groupmorph}, Bi-JROS \cite{fan2024bi}, AC-DMiR \cite{khor2023anatomically}, PC-Reg \cite{yin2023pc}, PGCNet \cite{tan2023progressively}, and WS-BCNet \cite{jian2022weakly} use manually annotated segmentation labels to incorporate anatomical constraints in learning DVFs. 
Manual annotation is laborious for 3D medical image segmentation. To address this challenge, few-shot learning approaches have emerged as a promising solution. Methods such as PC-Reg-RT \cite{he2021few} and GL-Net \cite{ma2023deformable} generate pseudo-labels using segmentation networks pre-trained on limited labeled data, but they often exhibit insufficient precision for Regions of Interest (ROI) in complex anatomical cases.
Thus, \textbf{(Q1)}: How can we obtain effective representations of ROIs from images to facilitate downstream registration tasks? This remains a significant challenge.

\textbf{Unsupervised DMIR methods} typically process full image pairs without ground truth, as shown in Fig. \ref{img1}(b). Methods such as VM \cite{balakrishnan2019voxelmorph}, RDN \cite{hu2022recursive}, NICE-Net \cite{meng2022non} and SDHNet \cite{10042453} use CNNs to estimate DVF, then align moving images with fixed ones through spatial transformer networks (STN) \cite{jaderberg2015spatial}.
These methods demonstrate impressive speed and accuracy, but often suffer from misalignment and distortion by ignoring anatomical structures, ROI interactions, and irrelevant regions (IR). Other approaches like TransMorph \cite{chen2022transmorph}, NICE-Trans \cite{meng2023non}, ModeT \cite{wang2023modet}, XMorpher \cite{shi2022xmorpher}, and DMR \cite{chen2022deformer} use Transformer to model long-range spatial relationships. However, they similarly neglect ROI interference during alignment, causing registration errors.
Therefore, \textbf{(Q2)}: 
How can independent ROI alignment between image pairs be achieved in an unsupervised manner? This remains a critical challenge.

In summary, the motivation is twofold: (\textbf{Q1}) focuses on constructing effective representations of ROIs in images, and (\textbf{Q2}) aims to utilize these effective representations to achieve independent alignment of ROIs in downstream registration tasks through unsupervised learning.

In this study, we propose EASR-DCN, a new ROI-based framework for unsupervised DMIR (Fig. \ref{img1}c). The architecture comprises two core modules: 1) an Effective Anatomical Structure Representation (EASR) module using Gaussian Mixture Models (GMM) to extract ROI representations from voxel intensities, and 2) a Divide-and-Conquer Network (DCN) that learns ROI feature alignments through separate channels before integrating them into a comprehensive DVF.
The main contributions of our work are four-fold. 
\begin{itemize}
\item {\color{black} We propose a new EASR-DCN framework that leverages the corresponding ROI pairs to establish anatomical correspondence between two images. This framework integrates two synergistic components: (1) an EASR module that manages effective ROI representations, and (2) a DCN that performs independent ROI alignment.}

\item We introduce a new EASR strategy that divides images into multiple effective ROIs using GMM and investigates the optimal number of ROI representations.

\item We present a novel DCN that independently learns ROIs' feature alignments through distinct channels in the encoding stage and combines these correspondences in the decoding stage to generate a comprehensive DVF with reduced potential interference.

\item Extensive experiments demonstrate that our proposed EASR-DCN model surpasses existing unsupervised registration techniques and is competitive with weakly-supervised methods requiring labeled training images across four clinical applications.
\end{itemize}
\section{Related work}
\label{sec:Related}
\subsection{Representation Learning in Image Registration}
Representation learning is a fundamental problem in computer vision, and its core objective in DMIR is the effective representation of ROIs for precise and independent analysis. 

In recent years, it has garnered significant attention in the field of DMIR, where typical methods aim to represent the entire target image using various anatomical structures. For example, GL-Net \cite{ma2023deformable} and PC-Reg-RT \cite{he2021few} train segmentation networks for ROI representation using a few-shot learning strategy. However, due to the complexity of some anatomical features, it is often impractical to accurately delineate all anatomical structures based on a single segmentation network, leading to inaccuracies in downstream registration tasks.

To address this challenge, a strategy is proposed to represent images using multiple effective ROIs with distinct intensities. It should combine anatomical structures with similar voxel intensities into comprehensive ROIs and ensure segmentation accuracy by carefully managing the number of ROIs.

In this study, we thoroughly investigate the optimal number of ROI representations to ensure accurate ROI segmentation and provide valuable guidance for downstream registration.
\subsection{ROI-based Multiple-channel Alignment}
Alignment has been a central focus in DMIR for several decades, with techniques primarily emphasizing a global alignment strategy, such as VoxelMorph \cite{balakrishnan2019voxelmorph}, TransMorph \cite{chen2022transmorph}, XM \cite{shi2022xmorpher}, NICE-Net \cite{meng2022non}, DMR \cite{chen2022deformer}, and NICE-Trans \cite{meng2023non}. 
These methods ssume perfect voxel-wise matching between moving and target images but ignore ROI anatomy, often causing DVF folds/disorders and resulting image abnormalities.

The recent iterative SAMReg \cite{huang2024one} incorporates the SAM \cite{kirillov2023segment} segmentation model into an unsupervised iterative process to facilitate independent ROI alignment within the ROI-based registration framework. However, the iterative registration process is slow and does not meet real-time clinical requirements.

In this paper, we propose an ROI-based deep learning registration method that replaces global alignment with a DCN for independent ROI alignment between image pairs.
\begin{figure*}
\centering
\includegraphics[width=0.999\textwidth]{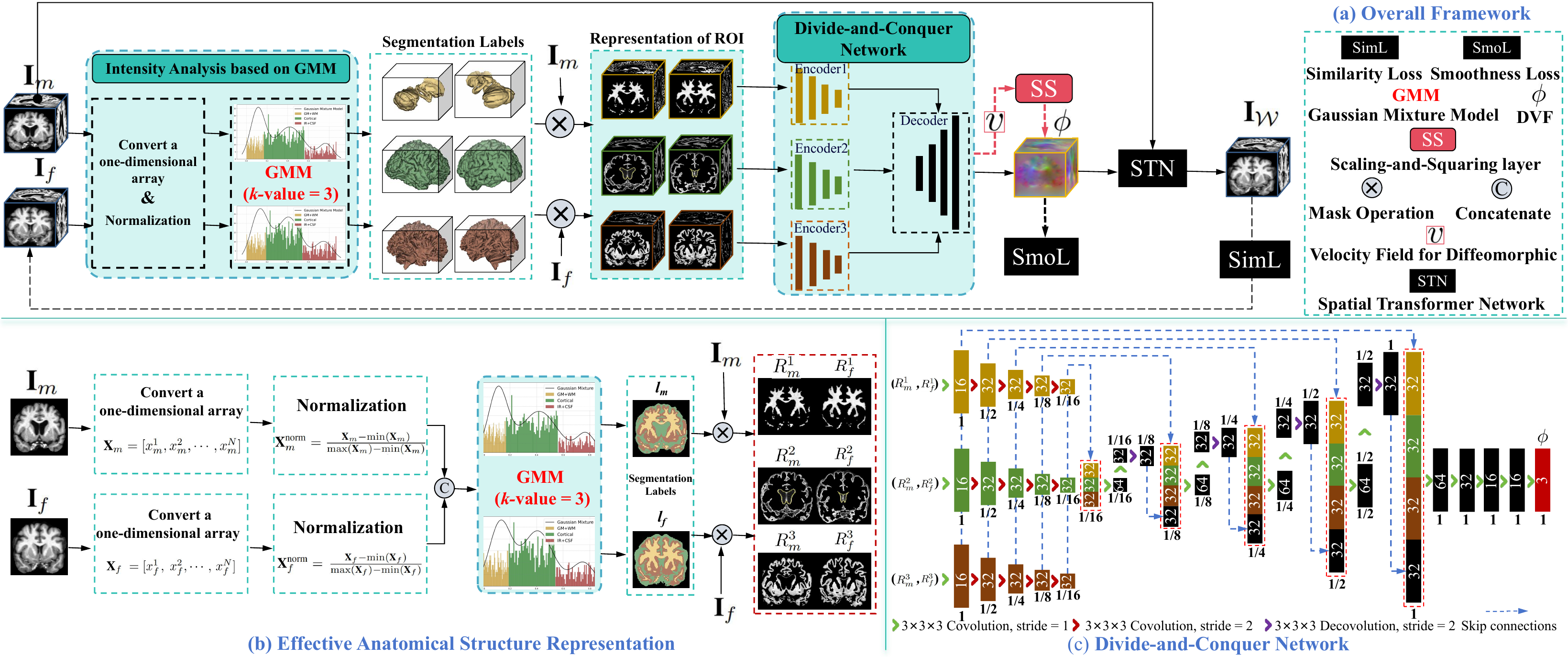}
\caption{Overall framework of our EASR-DCN model.
(a) EASR-DCN uses GMM with a specified $k$-value for intensity analysis, representing both moving ($\textbf{I}_m$) and fixed ($\textbf{I}_f$) images as multiple effective ROIs. These ROIs are then independently aligned using the DCN to produce a complete DVF ($\phi$), which is used to warp the $\textbf{I}_m$ to generate $\textbf{I}_{\mathcal{W}}$. The pink dotted line indicates the workflow for diffeomorphic registration. $v$ represents the velocity field for the diffeomorphic branch. The pink block named SS represents the scaling-and-squaring layer.
(b) $\textbf{I}_m$ and $\textbf{I}_f$ images are first converted into one-dimensional arrays. After regularization, these arrays are concatenated and analyzed by the GMM, with the $k$-value guiding accurate ROI representation.
(c) Paired ROIs are fed into the encoders of the DCN for independent feature alignment. The correspondences are then integrated to form a comprehensive DVF $\phi$.
}
\label{overall framework}
\end{figure*}
\begin{figure}
\centering
\includegraphics[width=0.49\textwidth]{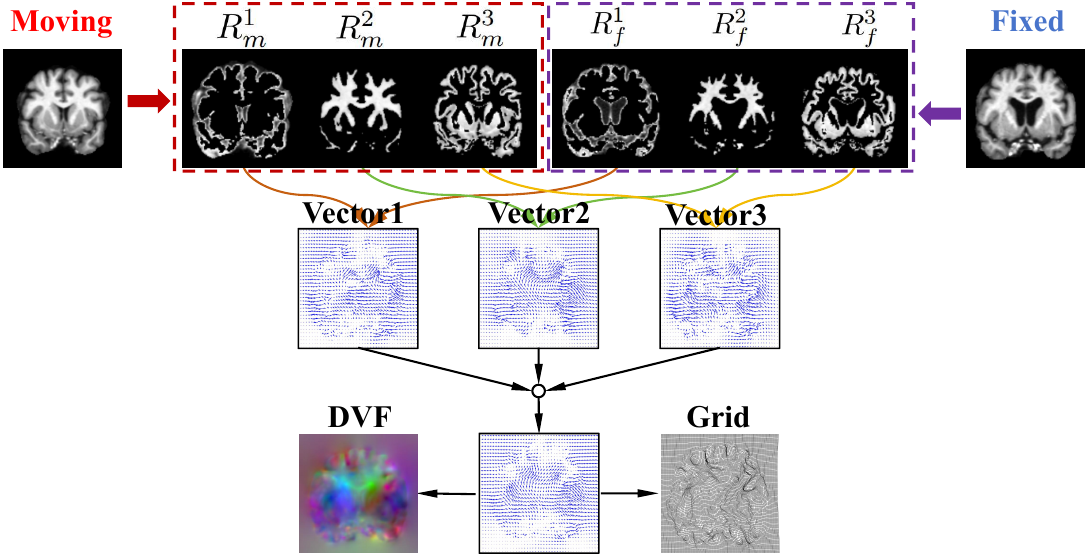}
\caption{Generation of a complete DVF on OASIS Brain images. The moving and fixed images are first represented by three valid ROIs, respectively. These ROIs are then fed into different encoders to generate three vector fields: Vector1, Vector2, and Vector3. Finally, these three vector fields are synthesized by a decoder to produce a complete DVF.}
\label{fusion}
\end{figure}
\section{Methods} 
\label{method}
\subsection{Preliminaries}
DMIR establishes voxel-level correspondences between the moving image ($\textbf{I}_m$) and the fixed image ($\textbf{I}_f$) through a spatial mapping $\phi(x) = x + \textbf{u}(x)$, where $x$ denotes a location within the domain $\rm{\Omega} \subset \mathbb{R}^{H \times W \times D}$, and $\textbf{u}(x)$ is the displacement fields. The mapping $\phi(x)$ is used to warp $\textbf{I}_m$ to align it with $\textbf{I}_f$, so that each voxel in $\textbf{I}_m$ corresponds to a voxel in the warped image $\textbf{I}_{\mathcal{W}}$ (denoted as $\textbf{I}_m \circ \phi$) through a STN \cite{jaderberg2015spatial}.

In unsupervised learning-based DMIR, a network $\mathcal{F}_{\theta}$ is trained to estimate the DVF $\phi$ from $\textbf{I}_m \circ \phi$ and $\textbf{I}_f$: $\phi = \mathcal{F}_{\theta} (\textbf{I}_m, \textbf{I}_f)$. The network parameters $\theta$ are optimized by minimizing a combined loss function to balance the similarity between $\textbf{I}_{\mathcal{W}}$ and $\textbf{I}_f$, and the smoothness of the DVF:
\begin{equation}
{\rm \hat{\phi}} = \underset{{\rm {\phi}}}{\rm{arg~min}}~{\mathcal{L}_{\rm SimL}(\textbf{I}_f,\textbf{I}_m\circ \rm{\phi}) + \alpha\mathcal{L}_{SmoL}(\rm{\phi})},
\label{eq1}
\end{equation} 
where $\mathcal{L}_{\text{SimL}}$ and $\mathcal{L}_{\text{SmoL}}$ are the image similarity and smoothness loss functions, respectively, $\alpha$ is a hyperparameter, and $\circ$ denotes composition operator.
\subsection{Effective Anatomical Structure Representation}
Given that $\textbf{I}_m$ and $\textbf{I}_f$ each contain $N$ voxels (Fig. \ref{overall framework}a), they are first converted to one-dimensional arrays $\textbf{X}_m$ and $\textbf{X}_f$:
\begin{equation}
\begin{aligned}
\begin{cases}
{\textbf{X}}_m= [x_m^1, x_m^2, \cdots, x_m^N], \\
{\textbf{X}}_f\hspace{0.1cm}= [x_f^1\hspace{0.03cm}, \hspace{0.1cm} x_f^2, \cdots, \hspace{0.06cm}x_f^N],
\end{cases}
\end{aligned}
\end{equation}
and then min-max normalized to $\textbf{X}_m^{\text{norm}}$ and $\textbf{X}_f^{\text{norm}}$:
\begin{equation}
\begin{aligned}
\begin{cases}
{\textbf{X}}_{m}^{\text{norm}} = \frac{\textbf{X}_{m} - \min(\textbf{X}_{m})}{\max(\textbf{X}_{m}) - \min(\textbf{X}_{m})}, ~x\in[1, N]\\
{\textbf{X}}_{f}^{\text{norm}} = \hspace{0.05cm}\frac{\textbf{X}_{f} - \min(\textbf{X}_{f})}{\max(\textbf{X}_{f}) - \min(\textbf{X}_{f})}\hspace{0.1cm}, ~x\in[1, N].
\end{cases}
\end{aligned}
\end{equation}

Here, $\textbf{X}_m^{\text{norm}}$ and $\textbf{X}_f^{\text{norm}}$ are normalized one-dimensional arrays, with $\min(\textbf{X})$ and $\max(\textbf{X})$ denoting the minimum and maximum values of the array $\textbf{X}$, respectively. 
{\color{black}In our EASR, the GMM performs joint segmentation by analyzing the voxel intensities of both images simultaneously, rather than processing each image independently.} 
The normalized arrays $\textbf{X}_m^{\text{norm}}$ and $\textbf{X}_f^{\text{norm}}$ are concatenated and input into a GMM for statistical voxel intensity analysis.

As shown in Fig. \ref{overall framework} (b), the GMM combines anatomical structures with similar voxel intensities into three ROIs on OASIS brain MRI: Cortex + Gray Matter (GM), White Matter (WM), and IR + CSF. We visualized voxel intensities along with the three Gaussian components, observing that each corresponds to an effective ROI. The probability density function of the GMM is given by:

\begin{equation}
p(x|\pi_k, \mu_k, \Sigma_k) = \sum_{k=1}^{K} \pi_k \mathcal{N}(x|\mu_k, \Sigma_k),
\label{eq2}
\end{equation}
where $\pi_k$ represents the weight of the $k^\text{th}$ Gaussian distribution, with $\sum_{k=1}^{K} \pi_k = 1$, and $\mathcal{N}(x | \mu_k, \Sigma_k)$ denotes the $k^\text{th}$ Gaussian distribution with mean $\mu_k$ and covariance $\Sigma_k$.

The GMM parameters are estimated using the EM algorithm \cite{em}, with the responsibility for each voxel $x_i$ calculated as:
\begin{equation}
\begin{aligned}
\gamma_{ik} = \frac{\pi_k \, \mathcal{N}(x_i | \mu_k, \Sigma_k)}{\sum_{j=1}^{K} \pi_j \, \mathcal{N}(x_i | \mu_j, \Sigma_j)}.
\end{aligned}
\end{equation}

The GMM parameters are estimated as:

\begin{equation}
\begin{aligned}
\pi_k = \frac{1}{N} \sum_{i=1}^{N} \gamma_{ik},
\end{aligned}
\end{equation}
\begin{equation}
\begin{aligned}
\mu_k = \frac{\sum_{i=1}^{N} \gamma_{ik} x_i}{\sum_{i=1}^{N} \gamma_{ik}},
\end{aligned}
\end{equation}
%
\begin{equation}
\begin{aligned}
\Sigma_k = \frac{\sum_{i=1}^{N} \gamma_{ik} (x_i - \mu_k)(x_i - \mu_k)^{\rm{T}}}{\sum_{i=1}^{N} \gamma_{ik}}.
\end{aligned}
\end{equation}

Once the GMM parameters are estimated, voxel cluster labels are assigned by selecting the Gaussian component with the highest responsibility, expressed as: $c_{ik} = \arg\max_k \gamma_{ik}$.

The segmentation labels are reshaped to their original size of $\text{H} \times \text{W} \times \text{D}$, which can be written as: 
\begin{equation}
\begin{cases}
l_m = \text{reshape}(c_{ik}[1:N], (\text{H,W,D})), \\
l_f \hspace{0.1cm} = \text{reshape}(c_{ik}[N+1:2N], (\text{H,W,D})).
\end{cases}
\end{equation}
The original images are masked using the moving label ($l_m$) and the fixed label ($l_f$) to obtain the three effective ROIs, which can be written as: 
\begin{equation}
\begin{cases}
R_m^i = \textbf{I}_m \otimes l_m^i, \text{ $ i \in [1,k] $ }, \\
R_f^i \hspace{0.1cm}= \textbf{I}_f \hspace{0.07cm} \otimes l_f^i\hspace{0.1cm}, \text{ $ i \in [1,k] $ }.
\end{cases}
\end{equation}
The $\otimes$ denotes the masking operation, $k$ represents the optimal number of ROIs, and $i$ indexes the $i$-th ROI in the image.
\subsection{Divide-and-Conquer Netwok}
The DCN performs multichannel feature alignment using extracted ROIs from both $\textbf{I}_m$ and $\textbf{I}_f$, as shown in Fig. \ref{overall framework}(c). The channel count dynamically adapts to each organ's ROIs.

For OASIS Brain MR images, we use a three-channel network architecture corresponding to three ROIs: Cortex+GM, WM, and IR+CSF. During the encoding stage, pairs of ROIs are inputted into three independent encoders to learn their alignment separately. In the encoding process, a $3 \times 3 \times 3$ convolutional layer (stride 1) is used to extract features, followed by a $3 \times 3 \times 3$ convolutional layer (stride 2) to compute low-level features and down-sample them. Each convolution is followed by a LeakyReLU layer with parameter 0.2.

In the decoding stage, the encoded features from the three channels are integrated via skip connections and then decoded. Specifically, two consecutive convolution layers are used: a $3 \times 3 \times 3$ convolutional layer (stride 1) for decoding, followed by a $3 \times 3 \times 3$ deconvolutional layer (stride 2) for upsampling and computing high-level features. This alternating process efficiently propagates encoded features directly to the layers responsible for registration, reducing interference from Irrelevant Regions (IR). At the end of the decoder, the $3 \times 3 \times 3$ convolution layers (stride 1) are used to generate the DVF ($\phi$). In this way, IR interference is reduced and mutual interference is effectively mitigated among different anatomical structures, enhancing registration accuracy.

Fig. \ref{fusion} illustrates the generation of DVF in OASIS Brain MR images using our EASR-DCN model. The moving and fixed images are represented by three effective ROIs, denoted as $R_m^1$, $R_m^2$, $R_m^3$ and $R_f^1$, $R_f^2$, and $R_f^3$, respectively. For these ROIs, three feature-aligned vector fields, \textit{i.e.}, Vector1, Vector2, and Vector3, are generated through three independent encoders. A decoder then integrates these vector fields to produce a complete $\phi$ with a deformed grid.
\subsection{Diffeomorphic Deformation}
To improve the continuity and smoothness of $\phi$, we propose a diffeomorphic variant, EASR-DCN-diff. To achieve this, we employ a stationary velocity field $v$ along with the Scaling-and-Squaring (SS) layer \cite{arsigny2006log} to derive the diffeomorphic deformation field. The velocity field $v$ satisfies the ordinary differential equation related to $\phi$, which can be expressed as:
\begin{equation}
\begin{aligned}
\frac{\partial \phi^{(t)}}{\partial t} = v (\phi^{(t)}),
\end{aligned}
\label{11}
\end{equation}
where $\phi^{(0)}$ represents the identity transformation. We integrate $v$ over the interval $t \in [0, 1]$ to acquire $\phi^{(1)}$.

Following in \cite{mok2020fast} and \cite{mok2020large}, we utilize SS \cite{arsigny2006log} as a numerical integration to solve Eq. \ref{11}.
Specifically, $v$ is scaled by $1/2^{\mathcal{T}}$ to generate the initial deformation field $\phi^{1/2^{\mathcal{T}}}$, given by
\begin{equation}
\begin{aligned}
\phi^{1/2^{\mathcal{T}}} = p + \frac{v(p)}{2^{\mathcal{T}}},
\end{aligned}
\end{equation}
where $p$ signifies the position map in the spatial domain, and $\mathcal{T}$ denotes the time step.

Subsequently, $\phi^{(1)}$ is generated recursively from $\phi^{1/2^{\mathcal{T}}}$ through the spatial transformation function. The recursive process can be formulated as: $\phi^{1/2^{t-1}} = \phi^{1/2^{t}} \circ \phi^{1/2^{t}}$.
The $\circ$ is a composition operation identical to that in \cite{mok2020fast}. Therefore, the deformation field at time 1 is $\phi^{1} = \phi^{1/2} \circ \phi^{1/2}$.
\subsection{Unsupervised End-to-End Learning}
As shown in Fig. \ref{overall framework} (a), our EASR-DCN model can be trained end-to-end by measuring the similarity \cite{balakrishnan2019voxelmorph} between $\textbf{I}_{\mathcal{W}}$ and $\textbf{I}_f$. The similarity loss function is defined as:

\begin{equation}
\begin{aligned}
{\rm{N}}&{\rm{CC}} ({\rm{I}}_f, {\rm{I}}_\mathcal{W}) 
\\&= \sum\limits_{{{{{x}}}}\in\Omega} \frac{\sum_{{{{{x}}}}_{i}}({\rm{I}}_f({{{{x}}}}_{i})-\bar{{\rm{I}}_f}({{{{x}}}}))({\rm{I}}_\mathcal{W}({{{{x}}}}_{i})-{\bar{{\rm{I}}_\mathcal{W}}}({{{{x}}}}))}{\sqrt{\sum_{{{{{x}}}}_{i}}({\rm{I}}_f({{{{x}}}}_{i})-\bar{{\rm{I}}_f}({{{{x}}}}))^{2}\sum_{{{x}}_{i}}({\rm{I}}_\mathcal{W}({{{{x}}}}_{i})-{\bar{{\rm{I}}_\mathcal{W}}}({{{{x}}}}))^{2}}},
\label{eq6} 
\end{aligned}
\end{equation}
where ${{x}}_{i}$ denotes a local region centered at ${{x}}$. A higher ${\rm NCC}$ value indicates better alignment. The image similarity loss function is given by $\mathcal{L}_{\text{SimL}}(\text{I}_f, \text{I}_{\mathcal{W}}) = - \text{NCC}(\text{I}_f, \text{I}_{\mathcal{W}})$. Minimizing $\mathcal{L}_{\text{SimL}}$ helps $\text{I}_{\mathcal{W}}$ to approximate $\text{I}_f$.

To ensure flexibility in spatial transformation, we impose $\rm{L}_2$ regularization on $\phi$ to guarantee smoothness. A diffusion regularizer function \cite{balakrishnan2019voxelmorph} applied to the spatial gradients of DVFs is used to obtain a smooth $\phi$:

\begin{equation}
\mathcal{L}_{\rm SmoL}(\phi) = \sum\limits_{{{{\bf{x}}}}\in\rm{\Omega}} \| \nabla {{{\bf{u}}}({\bf{x}})}\|^{2},
\label{eq6}
\end{equation}
where $\nabla \textbf{u}({\bf{x}})$ denotes the gradient calculation. The complete loss function combines $\mathcal{L}_{\text{SimL}}$ and $\mathcal{L}_{\text{SmoL}}$:

\begin{equation}
\begin{aligned}
\mathcal{L} ({\rm{I}}_f,{\rm{I}}_m,\phi) =  \mathcal{L}_{\rm SimL}{({\rm{I}}_f, {\rm{I}}_m \circ \phi)} + \alpha \mathcal{L}_{\rm SmoL}{(\phi)},
\label{eq6}
\end{aligned}
\end{equation}
where $\alpha$ is a hyper-parameter to balance the two loss functions.
%
\begin{table*}
\centering
\scriptsize
\setlength\tabcolsep{0.7 pt}
\caption{Quantitative evaluation results for three public MRI datasets.
DSC, $|{J_{\phi}}|{\leq 0}$ (\%), and HdDist95 are evaluated for ten {\underline{Unsupervised}} DMIR methods. 
The {\color{blue} {blue numbers}} denote the best scores, and the {\color{orange} orange numbers} indicate the second-best scores. Standard deviations are shown in parentheses. Folds are presented in e-notation (e.g., \(1\textit{e}{-2} = 0.01\)).
}
\begin{tabular}{|c|c|c|c|c|c|c|c|c|c|}
\hline
\rowcolor{gray! 5}{Dataset}&\multicolumn{3}{c|}{OASIS Brain MR}&\multicolumn{3}{c|}{Cardiac MR}&\multicolumn{3}{c|}{Hippocampus MR}\\
\hline
\rowcolor{gray! 5}{Metric} & DSC (\%) & $|{J}_{\phi}|{\leq 0}$ (\%) & HdDist95  & DSC (\%) & $|{J}_{\phi}|{\leq 0}$ (\%) & HdDist95& DSC (\%) & $|{J}_{\phi}|{\leq 0}$ (\%) & HdDist95 \\
\hline
Initial & $61.04~(5.11)$ & - & $3.975~(0.752)$ & $64.51~(11.83)$ &-&$5.938 ~ (0.827)$ & $60.92~(11.37)$ &-&$4.081 ~ (0.972)$\\
SyN (Baseline)\cite{avants2008symmetric} & $76.35~(2.73)$ &  \color{blue}{$0$} & $2.286~(0.374)$ & $74.85~ (10.60)$ & \color{blue}{$0.44e{-2}$} & $5.482 ~ (0.793)$& $71.14~(9.68)$ & \color{blue} ${1.26e{-6}}$ &  $3.269~(1.273)$\\
VM (TMI'2019) \cite{balakrishnan2019voxelmorph} & $78.89~(2.55)$& $5.57e{-2}$ & {{$2.107~(0.358)$}} & $75.69 ~(9.92)$ & $6.28e{-2}$ & $5.306 ~ (0.775)$& $74.29~(7.85)$ & $5.92e{-2}$ & $2.753~(1.219)$\\
TM (MedIA'2022)\cite{chen2022transmorph} & $80.68~(2.39)$ & $9.64e{-2}$ & $2.048~(0.350)$ & $77.05~(9.08)$ & $8.21e{-2}$ & $5.004~(0.718)$& $75.69~(7.09)$ & $7.30e{-2}$ & $2.625~(1.204)$\\
HM (IPMI’2021)\cite{hoopes2021hypermorph} & $79.70~(2.46)$ & $1.28e{-2}$ & $2.086~(0.355)$ & $76.49~(9.27)$ & $1.36e{-2}$ & $5.110~(0.739)$& $75.32~(7.41)$ & $1.50e{-2}$ & $2.697~(1.208)$\\
XM (MICCAI'2022)\cite{shi2022xmorpher} & $81.13~(2.44)$ & $7.59e{-2}$ & $2.004~(0.308)$ & $78.25~(8.59)$ & $7.40e{-2}$ & $4.803~(0.711)$& $ 74.60~(7.36)$ & $5.87e{-2}$ & $2.659~(1.210)$\\
CorrMLP {\color{black}(CVPR'2024)} \cite{meng2024correlation} & $87.13~(1.80)$ & $6.52e{-2}$ & {$1.958~(0.290)$} & $85.91~(6.52)$ & $5.90e{-2}$ & $3.827~(0.571)$& $78.54~(7.01)$ & $8.62e{-2}$ & $2.415~(1.190)$\\
NICE-Net (MICCAI'2022) \cite{meng2022non} & $83.59~(2.16)$ & $4.62e{-2}$ &$1.990~(0.304)$&$80.71~(7.94)$ & $5.56e{-2}$ & $4.725~(0.602)$& $76.49~(7.33)$ & $4.09e{-2}$ & $2.617~(1.205)$\\
%
DMR (MICCAI'2022) \cite{chen2022deformer} & $83.71~(2.10)$ & $5.17e{-2}$ & $2.008~(0.299)$ &$80.93~(7.90)$ & $5.83e{-2}$ & $4.281~(0.610)$& $76.57~(7.28)$ & $6.24e{-2}$ & $2.601~(1.204)$\\
SDHNet  (TMI'2023)\cite{10042453} & $85.19~(1.92)$ & $7.17e{-2}$ &$1.995~(0.293)$&$83.17~ (7.08)$ & $6.38e{-2}$ & $3.997~(0.592)$& $76.92~(7.13)$ & $2.47e{-2}$ & $2.579~(1.206)$\\
NICE-Trans (MICCAI'2023)\cite{meng2023non} & $85.46~(1.96)$ & $6.83e{-2}$ &$1.972~(0.292)$&$83.75~(6.83)$ & $6.24e{-2}$ & $3.980~ (0.584)$& $77.03~(7.12)$ & $5.13e{-2}$ & $2.499~(1.195)$\\
 EASR-DCN (Ours) & \color{blue}{$89.20~(1.77)$}& $1.50e{-2}$ &\color{blue}${1.924~(0.281)}$& \color{blue}${88.70~(5.99)}$ & $1.06e{-2}$ & \color{blue}${3.597~(0.550)}$ & \color{blue}{$80.04~(6.94)$} & $1.62e{-2}$ & \color{blue}{$2.397~(1.186)$}\\
EASR-DCN-diff (Ours) & {\color{orange} {$88.04~(1.84)$}} & {\color{orange}$1.19e{-2}$} &{\color{orange}$1.944~(0.288)$}& {\color{orange}$86.94~(6.15)$} & {\color{orange}$0.89e{-2}$} & {\color{orange}$3.826~(0.568)$}& {\color{orange}$78.62~(7.03)$} & {\color{orange}$0.55e{-2}$} & {\color{orange}$2.408~(1.191)$}\\
\hline 
\end{tabular}
\label{unsupervised}
\end{table*}
\begin{table*}
\centering
\scriptsize
\setlength\tabcolsep{0.8 pt}
\caption{Quantitative evaluation results for three public MRI datasets.
DSC, $|{J_{\phi}}|{\leq 0}$ (\%), and HdDist95  are evaluated for ten {\underline{Weakly-supervised}} DMIR methods. 
The {\color{blue} {blue numbers}} denote the best scores, and the {\color{orange} orange numbers} indicate the second-best scores. Standard deviations are shown in parentheses. Folds are presented in e-notation (e.g., \(1\textit{e}{-2} = 0.01\)).
}
\begin{tabular}{|c|c|c|c|c|c|c|c|c|c|}
\hline
\rowcolor{gray! 10}{Dataset}&\multicolumn{3}{c|}{OASIS Brain MRI}&\multicolumn{3}{c|}{Cardiac MRI}&\multicolumn{3}{c|}{Hippocampus MRI}\\
\hline
\rowcolor{gray! 10}{Metric} & DSC (\%) & $|{J}_{\phi}|{\leq 0}$ (\%) & HdDist95  & DSC (\%) & $|{J}_{\phi}|{\leq 0}$ (\%) & HdDist95& DSC (\%) & $|{J}_{\phi}|{\leq 0}$ (\%) & HdDist95 \\
\hline
Initial & $61.04~(5.11)$ & - & $3.975~(0.752)$ & $64.51~(11.83)$ &-&$5.938 ~ (0.827)$ & $60.92~(11.37)$ &-&$4.081 ~ (0.972)$\\
GL-Net (CMIG'2023) \cite{ma2023deformable} & $87.85~(1.98)$&$4.37e{-2}$& $1.902~(0.301)$ &$80.49~(7.73)$& $4.72e{-2}$ & $4.750~(0.677)$& $78.75~(6.72)$ & $4.89e{-2}$ & $2.496~(1.175)$\\
{PC-Reg} (MedIA'2023) \cite{yin2023pc} & $87.94~(1.95)$& $5.28e{-2}$ & $1.944~(0.275)$ &$80.97~(7.70)$& $5.94e{-2}$ & $4.749~(0.672)$& $78.99~(6.69)$ & $5.03e{-2}$ & $2.494~(1.173)$\\
WS-VM (TMI'2019)\cite{balakrishnan2019voxelmorph} & $82.91~ (2.30)$& $4.66e{-2}$ & $2.015~(0.317)$  & $ 77.29~(8.97)$& $5.31e{-2}$ & $4.995~(0.707)$& $76.19~(7.14)$ & $4.55e{-2}$ & $2.706~(1.206)$\\
PC-Reg-RT (JBHI'2021)  \cite{he2021few} & $86.95~ (2.08)$& $2.74e{-2}$ & $1.971~(0.292)$ &$80.06~(7.82)$& $2.99e{-2}$ & $4.760~(0.682)$& $78.31~(6.84)$ & $2.79e{-2}$ & $2.570~(1.197)$\\
RsegNet (TASE'2021) \cite{qiu2021rsegnet} & $85.03~(2.17)$& {$0.75e{-2}$} &$1.998~(0.284)$ & $ 78.97~(8.05)$& $1.33e{-2}$ & $4.792~(0.701)$& $76.77~(7.19)$ & \color{orange}{$1.42e{-2}$} & $2.601~(1.214)$\\
SUITS (MedIA'2021) \cite{blendowski2021weakly} & $85.92~(2.11)$& $7.51e{-2}$&$2.005~(0.288)$ & $ 79.18~(7.94)$& $8.27e{-2}$ & $4.780~(0.695)$& $77.23~(7.02)$ & $7.38e{-2}$ & $2.597~(1.212)$\\
WS-GM (TMI'2024) \cite{tan2024groupmorph} & \color{blue}{$90.91~(1.70)$}& $2.57e{-2}$ &\color{blue} {$1.915~(0.273)$}& \color{blue} $89.15~(5.81)$& $3.89e{-2}$ & \color{blue} $3.566~(0.544)$ & \color{orange} $80.02~(6.98)$ & $2.19e{-2}$ & \color{orange} $2.399~(1.190)$\\
Bi-JROS (CVPR'2024)\cite{fan2024bi} & $88.78~(1.74)$& $3.49e{-2}$ & $1.953~(0.288) $ &$82.06~(6.55)$& $3.58e{-2}$ & $3.996~(0.607)$& $79.55~(6.53)$ & $3.86e{-2}$ & $2.487~(1.195)$\\
AC-DMiR (MedIA'2023)\cite{khor2023anatomically} & $86.20~(2.19)$& $4.48e{-2}$&$1.964~(0.290) $& $ 79.44~(7.90)$& $4.96e{-2}$ & $4.772~(0.691)$ & $77.67~(6.94)$ & $4.66e{-2}$ & $2.590~(1.206)$\\
PGCNet (MICCAI'2023)\cite{tan2023progressively} & $87.03~(2.06)$& $4.97e{-2}$&$1.952~(0.283) $&$80.15~(7.79)$& $2.94e{-2}$ & $4.753~(0.680)$& $78.57~(6.80)$ & $5.22e{-2}$ & $2.557~(1.190)$\\
WS-BCNet (MICCAI'2023)\cite{jian2022weakly} & $86.33~(2.14)$& $3.94e{-2}$&$1.947~(0.290)$&$79.62~(7.88)$& $3.72e{-2}$ & $4.766~(0.689)$& $77.90~(6.88)$ & $3.75e{-2}$ & $2.585~(1.201)$\\
 EASR-DCN (Ours) & {\color{orange}{$89.20~(1.77)$}}& {\color{orange}$1.50e{-2}$} & {\color{orange}${1.924~(0.281)}$} & {\color{orange}$88.70~(5.99)$} & {\color{orange}$1.06e{-2}$} & {\color{orange}$3.597~(0.550)$}& \color{blue}{$80.04~(6.94)$} & $1.62e{-2}$ & \color{blue}{$2.397~(1.186)$}\\
EASR-DCN-diff (Ours) & {$88.04~(1.84)$} & {\color{blue}$1.19e{-2}$} &{$1.944~(0.288)$}& {$86.94~(6.15)$} & {\color{blue}$0.89e{-2}$} & {$3.826~(0.568)$}& {$78.62~(7.03)$} & {\color{blue}$0.55e{-2}$} & $2.408~(1.191)$\\
\hline 
\end{tabular}
\label{weaklyunsupervised}
\end{table*}
%
\section{Experiments and Results}
\label{experiments}
\subsection{Data and Pre-processing}
We evaluated the proposed EASR-DCN model on three public MRI datasets (Hippocampus \cite{simpson2019large}, OASIS Brain \cite{hering2022learn2reg}, Cardiac MRI \cite{zhuang2016multi}) and a Cardiac CT dataset \cite{zhuang2016multi}. Our evaluation focused on three key aspects: (1) intricate brain structure registration, (2) hippocampus MR image alignment, and (3) precise cardiac MR/CT image registration.

{\textbf{OASIS Brain MRI Dataset:}} The OASIS Brain MRI Dataset \cite{hoopes2021hypermorph} is a widely used resource for evaluating registration methods, specifically the neuronal version \footnote{Available at \url{https://github.com/adalca/neurite}.}. 
It consists of 414 images, each pre-aligned to a standard template with a resolution of 160 $\times$ 192 $\times$ 224 voxels and a voxel spacing of 1 mm $\times$ 1 mm $\times$ 1 mm. For our experiments, we split the dataset into three subsets: 300 images for training, 14 for validation, and 100 for testing. This results in 89,700 (300 × 299) image pairs for training, 182 (14 $\times$ 13) image pairs for validation, and 9,900 (100 $\times$ 99) image pairs for testing. The dataset also includes 35 subcortical segmentation maps to evaluate the performance of the registration methods.

{\textbf{Hippocampus MRI Dataset:}} 
The Hippocampus MRI dataset \cite{simpson2019large} (Task 4 of L2R 2020 \footnote{Available at \url{https://learn2reg.grand-challenge.org/Datasets/}}) contains 390 T1-weighted scans (64 × 64 × 64 voxels,  1 mm × 1 mm × 1 mm resolution). Split into 273 training, 39 validation and 78 test scans, it provides 74,256 (273 × 272), 1,482 (39 × 38), and 6,006 (78 × 77) image pairs. All scans have anterior/posterior annotations.

{\textbf{Cardiac CT/MRI Dataset:}} 
The Cardiac CT/MRI dataset \footnote{Available at \url{https://zmiclab.github.io/zxh/0/mmwhs/}.} is sourced from the MM-WHS 2017 Challenge and includes 60 images: 20 labeled and 40 unlabeled.
The labeled set contains seven clinical ROIs: ascending aorta (AO), left atrial cavity (LA), left ventricular cavity (LV), left ventricular myocardium (Myo), pulmonary artery (PA), right atrial cavity (RA), and right ventricular cavity (RV).
{\color{black}For cardiac MR images, our standardized pipeline includes bias correction using N4ITK, affine alignment of ROIs, resampling to an isotropic resolution of 96 × 80 × 96 voxels, and min-max intensity normalization to [0,1]. Similarly, for cardiac CT images, we apply bias correction with N4ITK, affine ROI alignment, resampling to 96 × 80 × 96 voxels, and min-max intensity normalization. These steps ensure consistent input quality.}
For the unsupervised dataset, the 40 unlabeled images are used to generate 1,560 (40 $\times$ 39) training pairs. Five labeled images are designated for validation and 15 labeled images are used for testing, resulting in 210 (15 $\times$ 14) testing pairs.
The weakly-supervised dataset generates 156 training pairs (13 × 12) from 15 labeled images, with 2 for validation and 3 producing 20 test pairs (5 × 4).
\subsection{Compared Methods}
We compared our EASR-DCN and its diffeomorphic variant (EASR-DCN-diff) against several state-of-the-art (SOTA) registration methods, including 
(a) SyN \cite{avants2008symmetric}, a top-performing method among traditional nonlinear deformation techniques, implemented using the ANTs software package \cite{avants2009advanced} with a maximum iteration setting of (200,100,50),  
(b) VoxelMorph (VM) \cite{balakrishnan2019voxelmorph}, 
(c) TransMorph (TM) \cite{chen2022transmorph},  
(d) HyperMorph (HM) \cite{hoopes2021hypermorph}, 
(e) XMorpher (XM) \cite{shi2022xmorpher}, 
(f) CorrMLP \cite{meng2024correlation}, 
(g) NICE-Net \cite{meng2022non}, 
(h) DMR \cite{chen2022deformer}, 
(i) SDHNet \cite{10042453}, 
(j) NICE-Trans \cite{meng2023non}, 
(k) GL-Net \cite{ma2023deformable}, 
(l) PC-Reg \cite{yin2023pc}, 
(m) WS-VM \cite{balakrishnan2019voxelmorph}, 
(n) PC-Reg-RT  \cite{he2021few}, 
(u) RsegNet \cite{qiu2021rsegnet}, 
(v) SUITS \cite{blendowski2021weakly}, 
(w) WS-GM \cite{tan2024groupmorph}, 
(r) Bi-JROS \cite{fan2024bi}, 
(s) AC-DMiR \cite{khor2023anatomically}, 
(t) PGCNet \cite{tan2023progressively}, and 
(x) WS-BCNet \cite{jian2022weakly}.
Among these competing methods, (a) is a traditional registration method, (b) to (j) are unsupervised learning-based registration methods, and (k) to (x) are supervised learning-based registration methods. For a fair comparison, all deep learning models (b)-(x) were retrained using their public implementations, with adjusted training parameters to achieve optimal registration performance.
\subsection{Implementation Details}
The EASR-DCN model was implemented using PyTorch \cite{paszke2017automatic} on an NVIDIA RTX3090 GPU with 24 GB of memory. We employed the Adam optimization with a learning rate of $1 \times 10^{-4}$ and a batch size of 1. The training was conducted with 200,000 iterations on the OASIS Brain and Hippocampus MRI datasets and 60,000 iterations on the Cardiac MRI/CT dataset. Validation was performed every 1,000 iterations for the OASIS Brain and Hippocampus MRI datasets, and every 300 iterations for the Cardiac MRI/CT dataset. The model with the highest validation score was selected for final testing.
\subsection{{Evaluation Metrics}}
Registration performance was evaluated using the Dice score (DSC) \cite{dice1945measures}, the percentage of voxels with non-positive Jacobian determinant ($|{{J}}_{\rm{\phi}}|{\leq 0}$) \cite{mok2020large}, and the 95\% maximum Hausdorff distance (HdDist95) \cite{huttenlocher1993comparing}. A better registration is indicated by a higher DSC, and lower HdDist95 and $|{{J}}_{\rm{\phi}}|{\leq 0}$.
\subsection{Results and Analysis}
We first compared the proposed EASR-DCN with a traditional iterative method and nine SOTA unsupervised learning-based methods to assess its effectiveness.

Table \ref{unsupervised} shows the registration results of the OASIS Brain MRI, Hippocampus MRI, and Cardiac MRI datasets, including initial affine normalization and results from SyN \cite{avants2008symmetric}, VM \cite{balakrishnan2019voxelmorph}, TM \cite{chen2022transmorph}, HM \cite{hoopes2021hypermorph}, XM \cite{shi2022xmorpher}, CorrMLP \cite{meng2024correlation}, NICE-Net \cite{meng2022non}, DMR \cite{chen2022deformer}, SDHNet \cite{10042453}, NICE-Trans \cite{meng2023non}, our EASR-DCN, and our EASR-DCN-diff.
\begin{figure}
\centering
\includegraphics[width=0.49\textwidth]{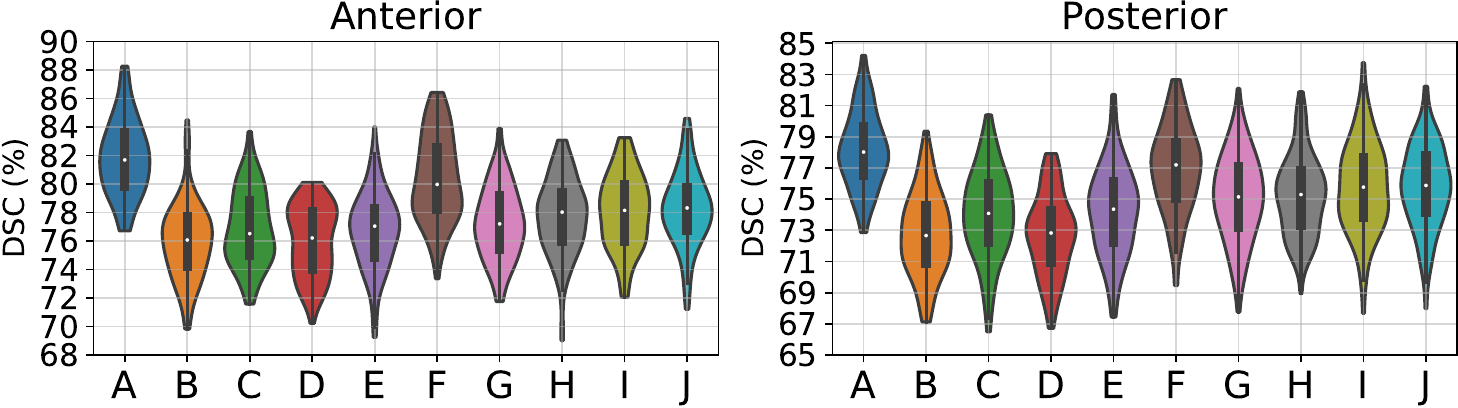}
\caption{Violin plots showing DSC of two hippocampal structures. A-J: our EASR-DCN, VM\cite{balakrishnan2019voxelmorph}, TM\cite{chen2022transmorph}, XM\cite{shi2022xmorpher}, HM \cite{hoopes2021hypermorph}, CorrMLP \cite{meng2024correlation}, NICE-Net \cite{meng2022non}, DMR \cite{chen2022deformer}, SDHNet \cite{10042453}, and NICE-Trans \cite{meng2023non}.}
\label{subplot2}
\end{figure}
\begin{figure*}
\centering
\includegraphics[width=0.96\textwidth]{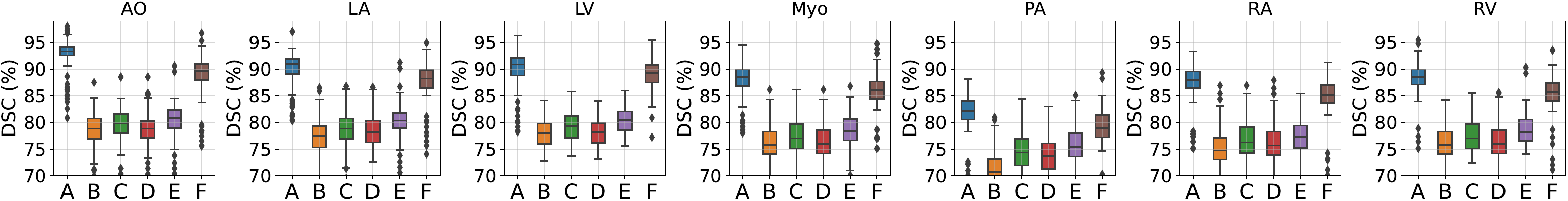}
\caption{Boxplot showing DSC of seven structures on Cardiac MRI. A-F: our EASR-DCN, VM\cite{balakrishnan2019voxelmorph}, TM\cite{chen2022transmorph}, XM\cite{shi2022xmorpher}, HM \cite{hoopes2021hypermorph}, and CorrMLP \cite{meng2024correlation}.}
\label{plot7}
\end{figure*}
\begin{figure*}
\centering
\includegraphics[width=0.96\textwidth]{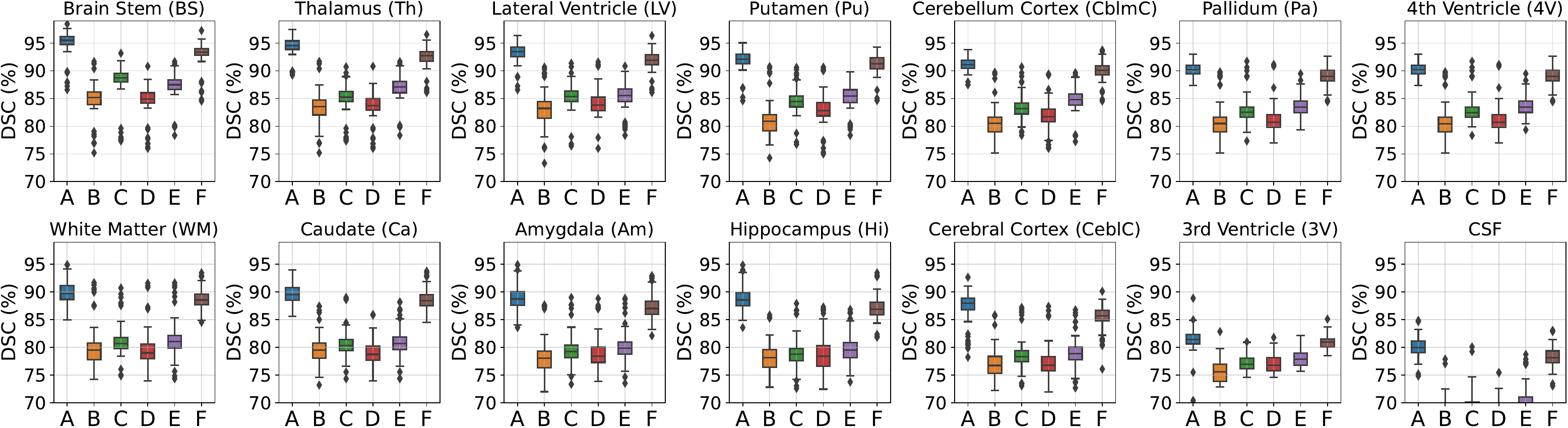}
\caption{Boxplot showing DSC of fourteen structures on Brain MRI. A-F: our EASR-DCN, VM\cite{balakrishnan2019voxelmorph}, TM\cite{chen2022transmorph}, XM\cite{shi2022xmorpher}, HM \cite{hoopes2021hypermorph}, and CorrMLP \cite{meng2024correlation}.}
\label{plot14}
\end{figure*}
\begin{figure}
\centering
\includegraphics[width=0.46\textwidth]{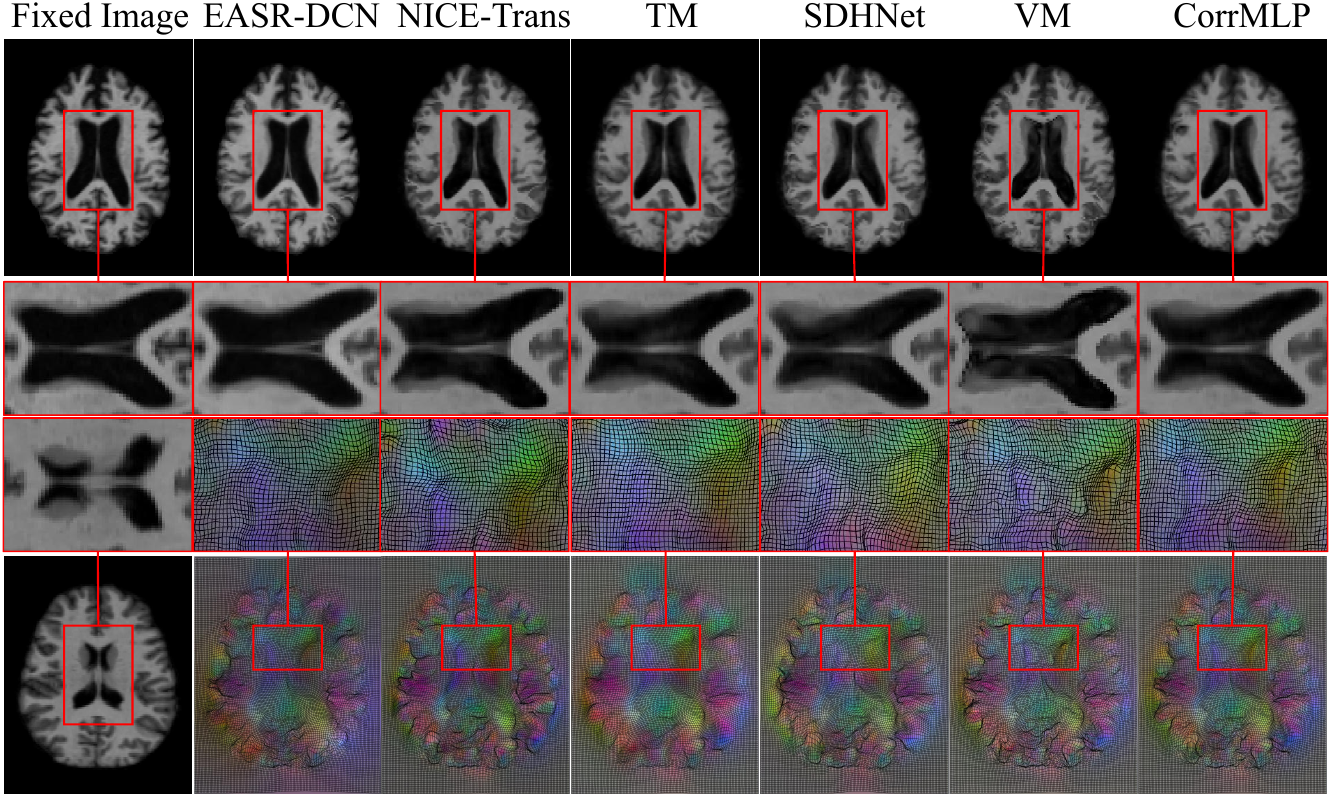}
\includegraphics[width=0.46\textwidth]{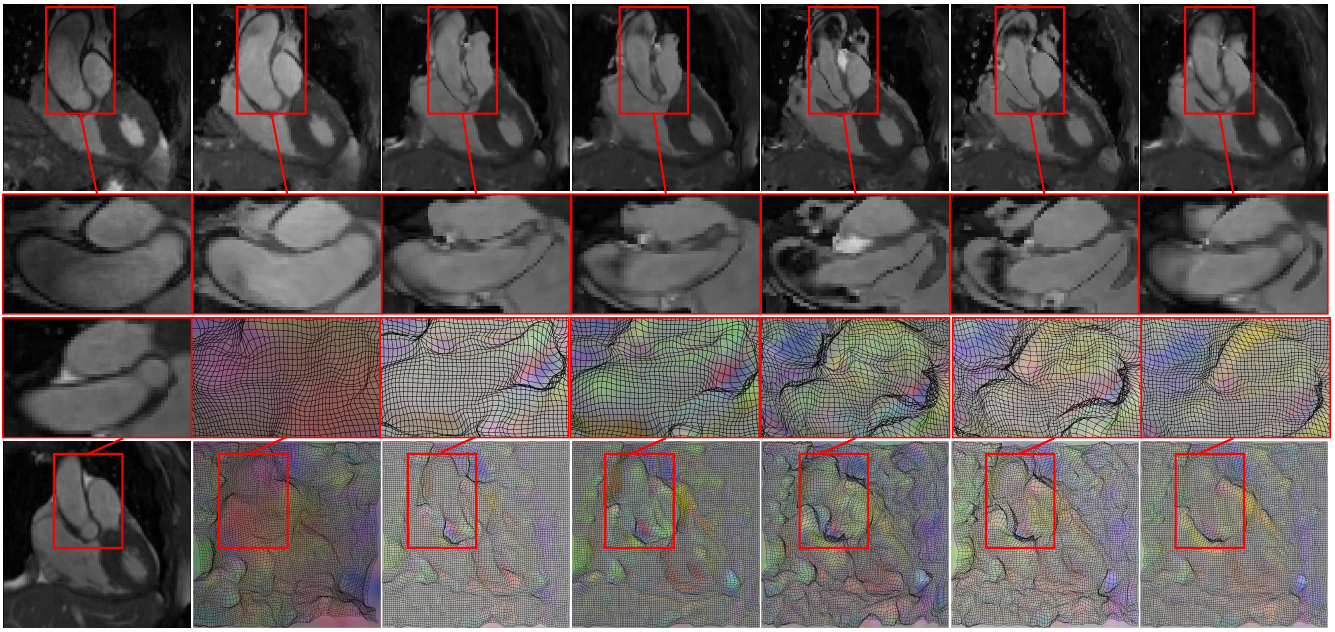}
\includegraphics[width=0.46\textwidth]{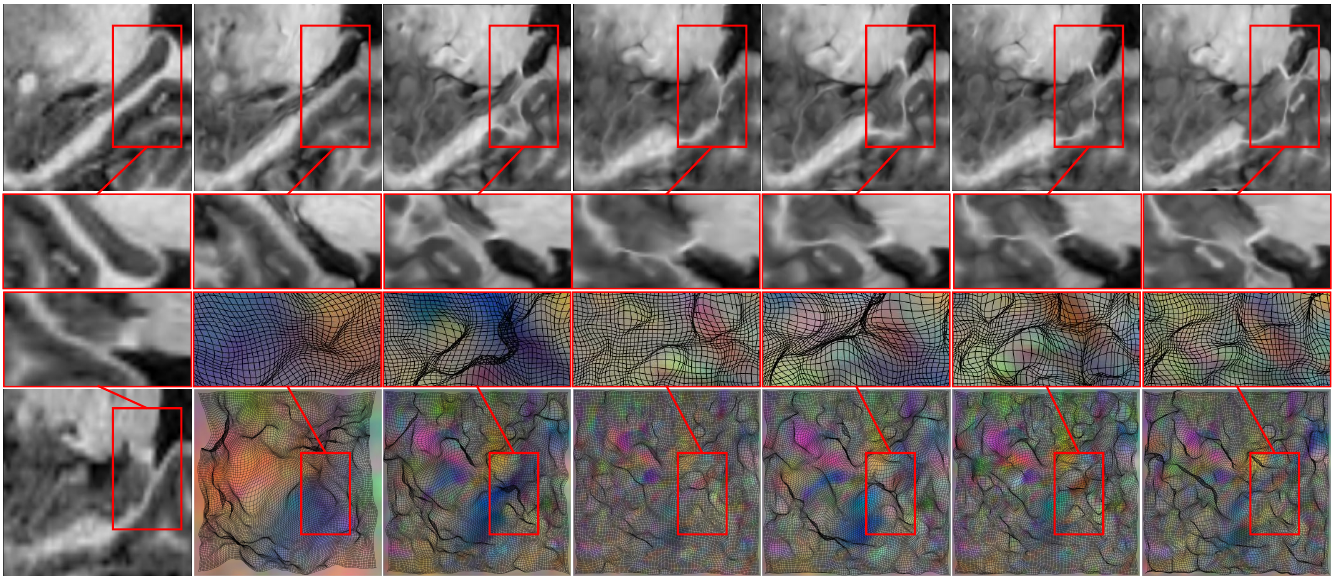}
\caption{Representative registration results from three MRI datasets: OASIS Brain MRI, Cardiac MRI, and Hippocampus MRI. Column 1 shows fixed images in odd rows and moving images in even rows. Columns 2-7 display the registration results of six unsupervised methods: our EASR-DCN, NICE-Trans \cite{meng2023non}, TM \cite{chen2022transmorph}, SDHNet \cite{10042453}, VM \cite{balakrishnan2019voxelmorph}, and CorrMLP \cite{meng2024correlation}, in order. Odd rows show warped images, while even rows show DVFs. The red rectangular box suggests a comparison of registration results.}
\label{demoreg}
\end{figure}
\begin{figure}
\centering
\includegraphics[width=0.47\textwidth]{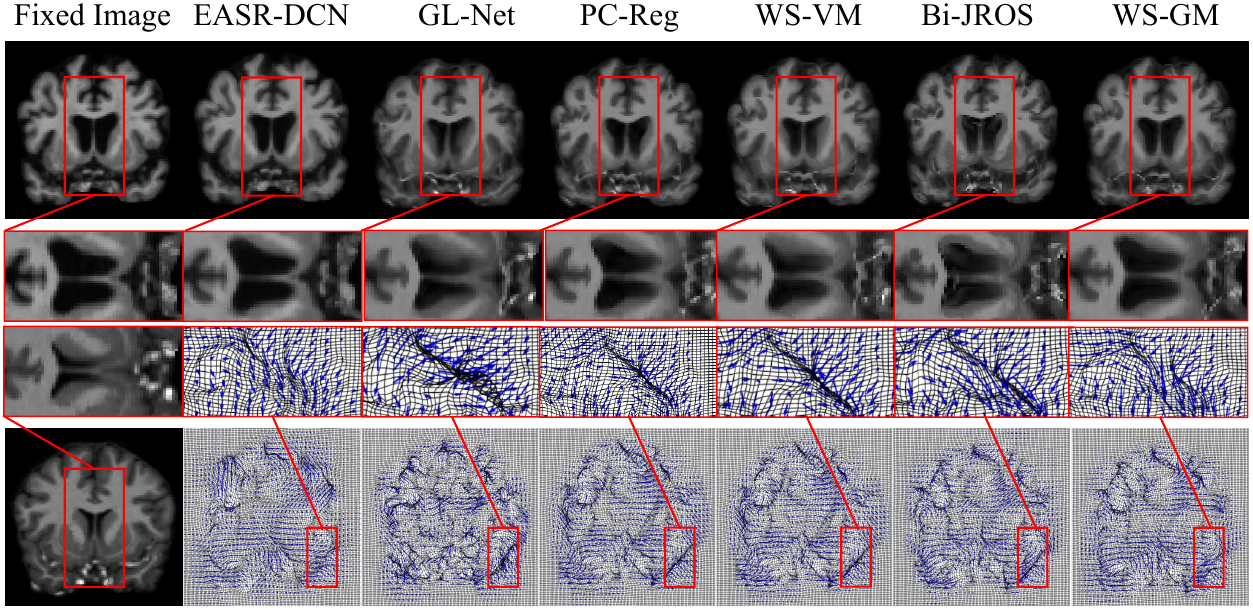}
\includegraphics[width=0.47\textwidth]{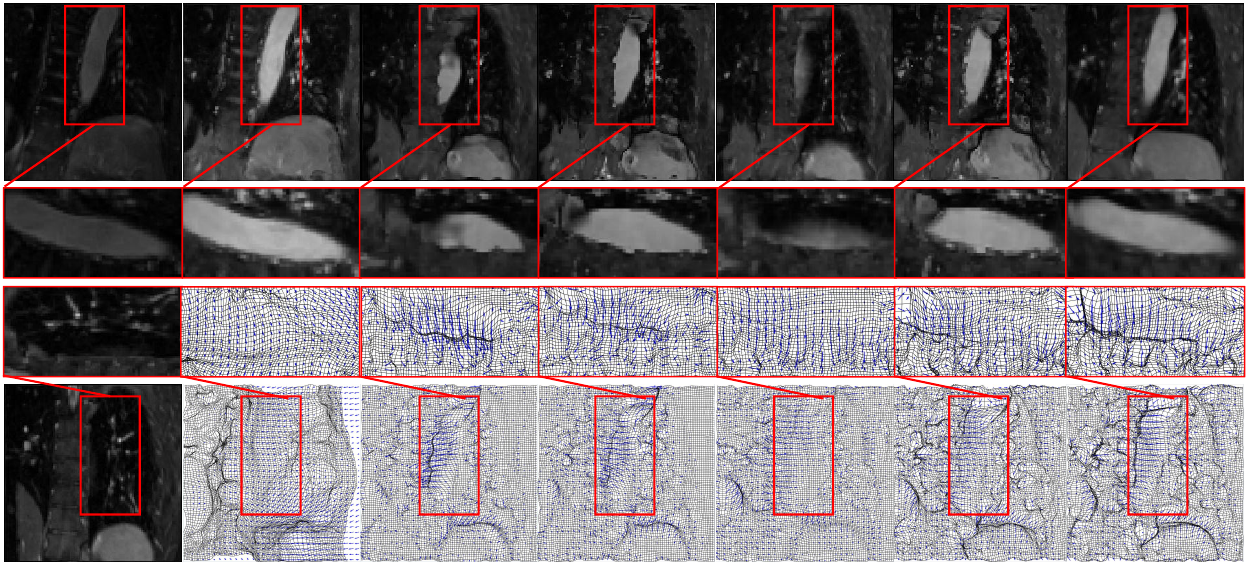}
\includegraphics[width=0.47\textwidth]{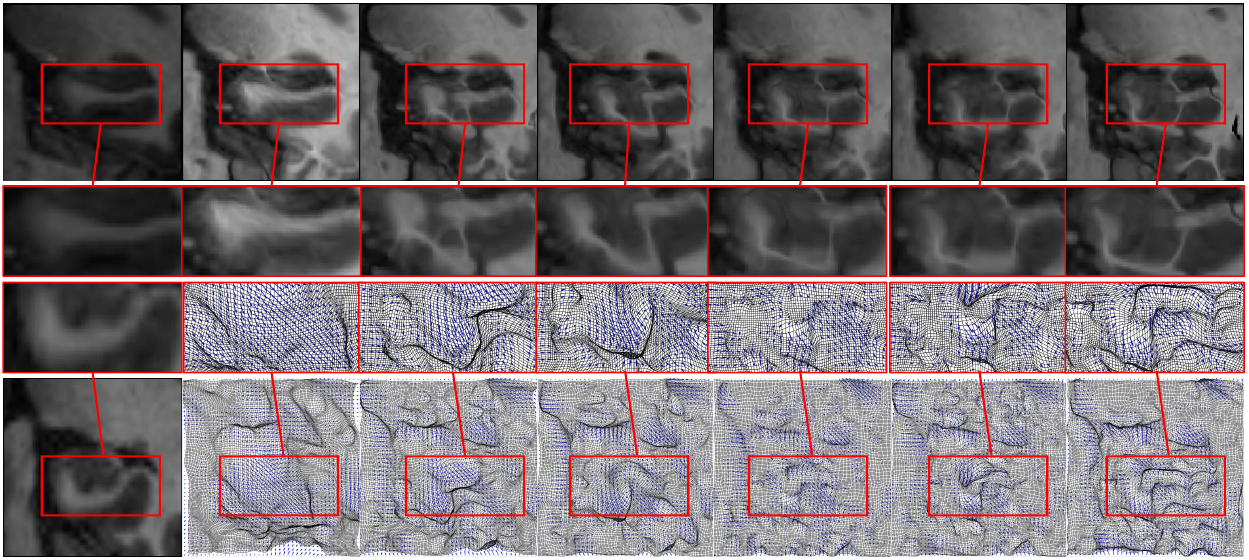}
\caption{Representative registration results from three MR datasets: OASIS Brain MRI, Cardiac MRI, and Hippocampus MRI. Column 1 shows fixed images in odd rows and moving images in even rows. Columns 2-7 display the registration results from our EASR-DCN and five weakly supervised methods: GL-Net \cite{ma2023deformable}, PC-Reg \cite{yin2023pc}, WS-VM \cite{balakrishnan2019voxelmorph}, Bi-JROS \cite{fan2024bi}, and WS-GM \cite{tan2024groupmorph}, in order. Odd rows show warped images, while even rows show DVFs. The blue arrows represent the displacement vector fields with the deformed grid. The red rectangular box suggests a comparison of registration results.}
\label{demo-weakly-reg}
\end{figure}
Our EASR-DCN model consistently outperforms other registration methods in most performance metrics. Specifically, EASR-DCN improves the average DSC by 2.07\% on the OASIS Brain MRI dataset, 2.79\% on the Cardiac MRI dataset, and 1.50\% on the Hippocampus MRI dataset, compared to the second-best method, CorrMLP \cite{meng2024correlation}. It also reduces HdDist95 by 0.034 on OASIS Brain MRI, 0.230 on Hippocampus MRI, and 0.018 on Cardiac MRI. Additionally, EASR-DCN decreases $|{J}_\phi|{\leq 0}$ by 0.0502 on OASIS Brain MRI, 0.0700 on Hippocampus MRI, and 0.0484 on Cardiac MRI, indicating fewer folds in DVFs compared to other learning-based methods.

In addition, we further reported the registration results of EASR-DCN and eleven SOTA weakly-supervised learning-based methods in Table \ref{weaklyunsupervised} including GL-Net \cite{ma2023deformable}, PC-Reg \cite{yin2023pc}, WS-VM \cite{balakrishnan2019voxelmorph}, PC-Reg-RT \cite{he2021few}, RsegNet \cite{qiu2021rsegnet}, SUITS \cite{blendowski2021weakly}, WS-GM \cite{tan2024groupmorph}, Bi-JROS \cite{fan2024bi}, AC-DMiR \cite{khor2023anatomically}, PGCNet \cite{tan2023progressively}, WS-BCNet \cite{jian2022weakly}, as well as our EASR-DCN and our EASR-DCN-diff.
Our EASR-DCN shows competitive registration accuracy compared to the top two weakly-supervised methods.
Particularly, compared to the best method, WS-GM \cite{tan2024groupmorph}, EASR-DCN improves the average DSC by 0.02\% and reduces HdDist95 by 0.002 on the Hippocampus MRI dataset. Meanwhile, EASR-DCN reduces $|{J}_\phi|{\leq 0} (\%)$ by 0.0107 on Brain MRI, 0.0283 on Cardiac MRI, and 0.0057 on Hippocampus MRI, demonstrating its superior capability in reducing folds.

Fig. \ref{subplot2} (violin plot) and Figs. \ref{plot7}–\ref{plot14} (boxplots) show the average DSC across Hippocampus, Cardiac, and OASIS Brain MRI datasets. For Hippocampus MRI, violin plots compare EASR-DCN with nine SOTA unsupervised methods at the anterior and posterior ends of the hippocampus. Cardiac MRI boxplots cover seven heart structures, while OASIS Brain MRI includes fourteen. EASR-DCN achieves the highest registration accuracy across all structures in all three datasets.

Figs. \ref{demoreg} and \ref{demo-weakly-reg} illustrate the registration results of the three MRI datasets. Rows 1 and 2 show the results for the OASIS Brain MRI dataset, rows 3 and 4 show the results for the Cardiac MRI dataset, and rows 5 and 6 show the results for the Hippocampus MRI dataset, respectively. The red rectangular box suggests a comparison of the registration results.

In Fig. \ref{demoreg}, the first column presents the moving and fixed images. Columns 2 through 7 show the results of six unsupervised learning-based registration methods: our EASR-DCN, NICE-Trans \cite{meng2023non}, TM \cite{chen2022transmorph}, SDHNet \cite{10042453}, VM \cite{balakrishnan2019voxelmorph}, and CorrMLP \cite{meng2024correlation}. In Fig. \ref{demo-weakly-reg}, the first column presents the initial moving and fixed images, while columns 2 through 7 display results from our EASR-DCN and five weakly-supervised learning-based registration methods: GL-Net \cite{ma2023deformable}, PC-Reg \cite{yin2023pc}, WS-VM \cite{balakrishnan2019voxelmorph}, Bi-JROS \cite{fan2024bi}, and WS-GM \cite{tan2024groupmorph}. 
It can be observed that our EASR-DCN shows superior registration results, mapping the moving images to fixed images more accurately than other methods. Besides, the even rows compare the folds in the DVFs, revealing that our EASR-DCN effectively eliminates folds.
%
%
\begin{table}
\centering
\scriptsize
\setlength\tabcolsep{2 pt}
\caption{Quantitative evaluation results for Cardiac CT.
DSC, $|{J_{\phi}}|{\leq 0}$ (\%), and HdDist95 are evaluated for ten {\underline{Unsupervised}} DMIR methods. 
The {\color{blue} {blue numbers}} denote the best scores, while the {\color{orange} orange numbers} indicate the second-best scores. Standard deviations are shown in parentheses. Folds are presented in e-notation (e.g., \(1\textit{e}{-2} = 0.01\)).
}
\begin{tabular}{|c|c|c|c|}
\hline
\rowcolor{gray! 5}{Dataset}&\multicolumn{3}{c|}{Cardiac CT}\\
\hline
\rowcolor{gray! 5}{Metric} & DSC (\%) & $|{J}_{\phi}|{\leq 0}$ (\%) & HdDist95   \\
\hline
Initial & $64.02~(12.53)$ & - & $6.248~(0.903)$\\
SyN (Baseline)\cite{avants2008symmetric} & $75.54~(10.73)$ &  \color{blue}{$0.513 e{-2}$} & $5.495~(0.878)$\\
VM (TMI'2019) \cite{balakrishnan2019voxelmorph} & $76.02~(10.54)$& $4.946 e{-2}$ & {{$5.261~(0.815)$}} \\
TM (MedIA'2022)\cite{chen2022transmorph} & $78.59~(10.43)$ & $5.527 e{-2}$ & $4.727~(0.806)$ \\
HM (IPMI’2021)\cite{hoopes2021hypermorph} & $77.63~(9.07)$ & $2.063 e{-2}$ & $4.958~(0.811)$\\
XM (MICCAI'2022)\cite{shi2022xmorpher} & $79.14~(8.02)$ & $5.226 e{-2}$ & $4.639~(0.794)$ \\
CorrMLP {\color{black}(CVPR'2024)} \cite{meng2024correlation} & $83.96~(4.24)$ & $4.538 e{-2}$ & {$3.709~(0.441)$}\\
NICE-Net (MICCAI'2022) \cite{meng2022non} & $80.26~(7.17)$ & $4.196 e{-2}$ &$4.447~(0.725)$\\
%
DMR (MICCAI'2022) \cite{chen2022deformer} & $80.79~(7.03)$ & $5.269 e{-2}$ &$4.406~(0.712)$\\
SDHNet  (TMI'2023)\cite{10042453} & $82.05~(5.84)$ & $4.829 e{-2}$ &$4.027~(0.695)$\\
NICE-Trans (MICCAI'2023)\cite{meng2023non} & $82.86~(5.29)$ & $4.746 e{-2}$ &$3.975~(0.583)$\\
 EASR-DCN (Ours) & \color{blue}{$85.72~(2.51)$}& $1.972 e{-2}$ &\color{blue}${2.944~(0.413)}$\\
EASR-DCN-diff (Ours) & {\color{orange} {$84.17~(3.18)$}} & {\color{orange}$1.009 e{-2}$} &{\color{orange}$3.660~(0.429)$}\\
\hline 
\end{tabular}
\label{CTregistration}
\end{table}
%
%
\begin{figure}
\center
\includegraphics[width=0.48\textwidth]{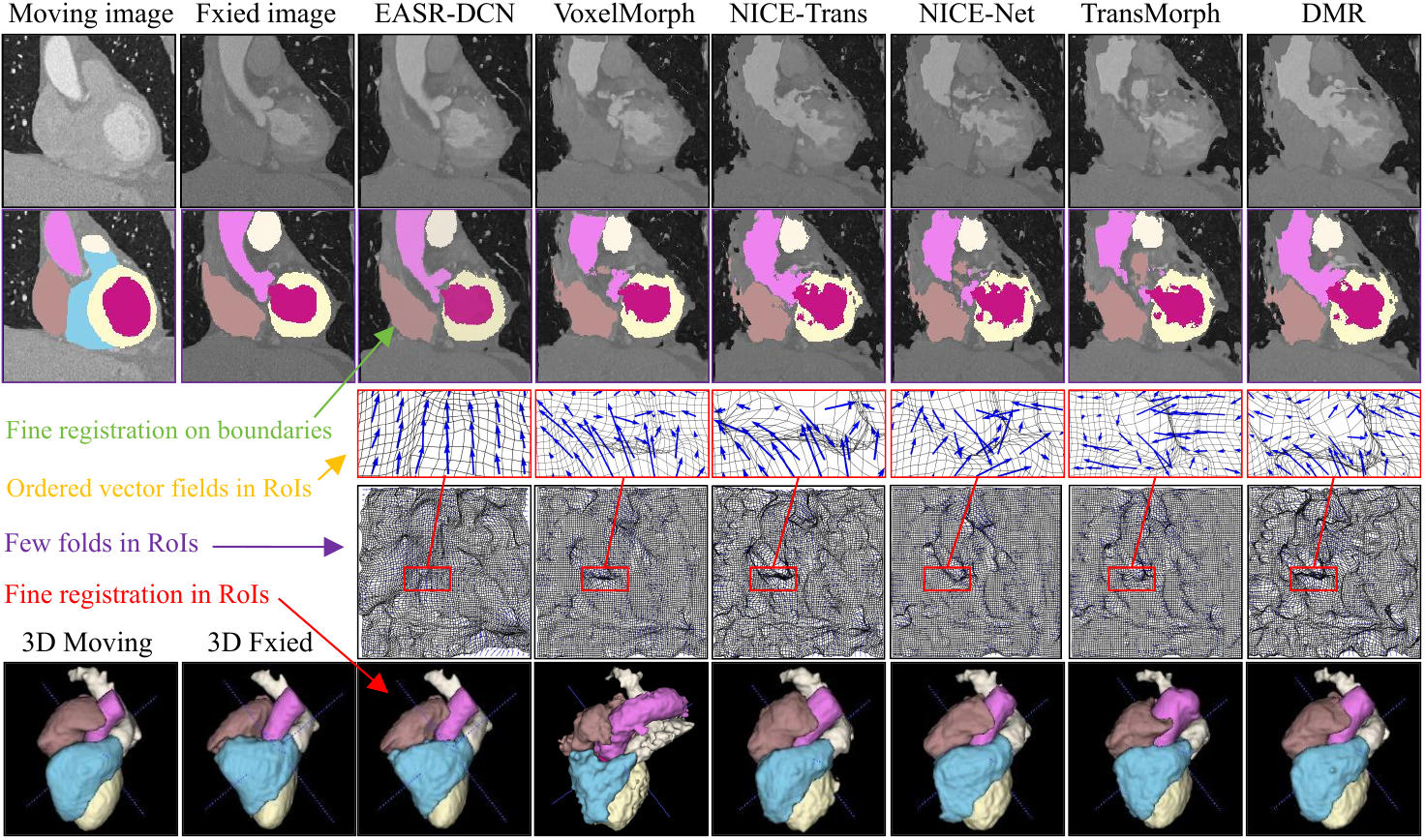}
\caption{Representative registration results from cardiac CT. Columns 1 and 2 show the moving image and the fixed image, respectively. Columns 3–8 display the registration results of six unsupervised methods: our EASR-DCN, VM \cite{balakrishnan2019voxelmorph}, NICE-Trans \cite{meng2023non}, NICE-Net \cite{meng2022non}, TM \cite{chen2022transmorph}, and DMR \cite{chen2022deformer}. Rows 1–4 in Columns 3–8 represent the warped image, warped image with warped labels, DVFs, and 3D deformation surface, respectively. The red rectangle highlights a comparison of the DVFs.}
\label{CTreg}
\end{figure}
\subsection{Cardiac CT Registration}
{\color{black}To further validate the applicability of our EASR-DCN in different modalities, we performed comprehensive registration experiments on cardiac CT images.

We compared our EASR-DCN with a traditional iterative method and nine unsupervised learning-based approaches. As shown in Table \ref{CTregistration}, our EASR-DCN consistently outperformed all competitors in the Cardiac CT dataset, including SyN \cite{avants2008symmetric}, VM \cite{balakrishnan2019voxelmorph}, TM \cite{chen2022transmorph}, HM \cite{hoopes2021hypermorph}, XM \cite{shi2022xmorpher}, CorrMLP \cite{meng2024correlation}, NICE-Net \cite{meng2022non}, DMR \cite{chen2022deformer}, SDHNet \cite{10042453} and NICE-Trans \cite{meng2023non}. Specifically, our EASR-DCN improved the average DSC by 1.76\% over the second-best method (CorrMLP), while also reducing HdDist95 by 0.765 and decreasing $|{J_{\phi}}|{\leq 0}$ (\%) by 0.026, indicating fewer folding artifacts in DVF.  

Fig. \ref{CTreg} presents results of cardiac CT registration. Columns 1 and 2 display the moving and fixed images, respectively. Columns 3 through 8 present the registration results from six unsupervised methods: our EASR-DCN, VM \cite{balakrishnan2019voxelmorph}, NICE-Trans \cite{meng2023non}, NICE-Net \cite{meng2022non}, TM \cite{chen2022transmorph}, and DMR \cite{chen2022deformer}. Within columns 3-8, rows 1-4 represent the moving image, moving image with labels, DVFs (grid + displacement fields), and 3D deformation surface, respectively.
Our EASR-DCN generates well-ordered displacement vector fields that effectively eliminate folds. Other methods show significant disorder of displacement vector fields at adjacent anatomical boundaries, demonstrating that our DCN module's independent establishment of ROI correspondences is effective. This successfully eliminates interference between anatomical regions, resulting in orderly displacement vector fields. Finally, the 3D surface registration results show that our EASR-DCN accurately aligns ROIs from moving to fixed images while maintaining structural integrity and smoothness. In contrast, other methods fail to effectively handle interference between anatomical regions during CT registration, leading to disordered vector fields that cause large-area registration errors and folded 3D surfaces.
}
%
%
\begin{table}
\center
\scriptsize
\setlength\tabcolsep{0.3 pt}
\caption{Quantitative evaluation results for brain MRI registration on the OASIS dataset from Learn2Reg 2021 Task 3. Validation results are from the challenge leaderboard, and test results are from the challenge organizers. {\color{blue}Blue numbers} indicate the best scores. Standard deviations are shown in parentheses.}
\begin{tabular}{|c|c|c|c|c|c|c|ccccccccc}
\hline
 ~ &\multicolumn{3}{c|}{Validation}&\multicolumn{3}{c|}{Test} 	\\
\hline
Method & DSC & HdDist95&SDlogJ&DSC&HdDist95&SDlogJ\\
\hline 
 Lv et al. \cite{lv2022joint} & $0.827 (0.013)$ & $1.722 (0.318)$ & $0.121 (0.015)$ & $0.80$ & $1.77$ & $0.08$  \\
%
 Siebert et al. \cite{siebert2021fast} & $0.846 (0.016)$ & $1.500 (0.304)$& $0.067 (0.005)$ & $0.81$  & $1.63$  & $0.07$  \\
 Mok et al. \cite{mok2021conditional}  & $0.861 (0.015)$  & $1.514 (0.337)$  &  $0.072 (0.007)$ & $0.82$  & $1.67$  & $0.07$  \\
 VM \cite{mok2020large}  & $0.861 (0.015)$  & $1.514 (0.337)$  &  $0.072 (0.007)$ & $0.82$  & $1.67$  & $0.07$  \\
TM \cite{chen2022transmorph} & $0.858 (0.014)$ & $1.494 (0.288)$& $0.118 (0.019)$ &$0.816$  & $1.692$ & $0.124$   \\
TM-Large \cite{chen2022transmorph} & $0.862 (0.014)$  & $1.431 (0.282)$ &$0.128 (0.021)$ & $0.820$ & $1.656$  & $0.124$ \\
Our EASR-DCN & {\color{blue}$0.908 (0.011)$}  & {\color{blue}$1.009 (0.195)$} &{\color{blue}$0.044 (0.008)$} & {\color{blue}$0.885$ }&{\color{blue} $1.298$}  & {\color{blue}$0.047$} \\
\hline
\end{tabular}
\label{L2R}
\end{table}
%
%
\begin{table}
\center
\scriptsize
\setlength\tabcolsep{2 pt}
\caption{Analysis of $k$-value for our EASR-DCN on the three  public MRI  datasets. The {\color{blue} {blue numbers}} denote the best scores.
}
\begin{tabular}{|c|c|c|c|c|c|c|ccccccccc}
\hline
 Dataset &\multicolumn{2}{c|}{OASIS Brain MRI}&\multicolumn{2}{c|}{Hippocampus MRI}&\multicolumn{2}{c|}{Cardiac MRI} 	\\
\hline
 $k$ & DSC (\%) & $|{J_\phi}|{\leq 0}$ (\%)&DSC (\%)&$|{J_\phi}|{\leq 0}$ (\%)&DSC (\%)&$|{J_\phi}|{\leq 0}$ (\%)\\
\hline 
 1 & $78.89$ & $5.57e{-2}$ & $74.29$ & $5.92e{-2}$ & $75.69 $ & $6.28e{-2}$  \\
%
 2  & $82.57$ & $3.29e{-2}$  & $77.53$ & $3.40e{-2}$ & $79.72$ & $4.39e{-2}$  \\
 3  & \color{blue}$89.20$ & \color{blue} $1.50e{-2}$ &  \color{blue} $80.04$ & \color{blue} $1.62e{-2}$ & $84.51 $ & $2.97e{-2}$  \\
 4 & $70.91$ & $9.73e{-2}$ &$65.79$ & $9.97e{-2}$& \color{blue} $88.70 $ & \color{blue} $1.06e{-2}$  \\
5 & $65.40$ & $1.19e{-1}$ &$52.69$ & $1.58e{-1}$& $60.38 $ & $1.27e{-1}$  \\
\hline
\end{tabular}
\label{kanalysis}
\end{table}
%
\subsection{Learn2Reg OASIS brain MRI registration}
{\color{black}To ensure a comprehensive evaluation, we follow the TransMorph experimental protocol \cite{chen2022transmorph} using the OASIS dataset from the MICCAI Learn2Reg 2021 Challenge (Task 3).

Table \ref{L2R} presents the quantitative results in both the validation set and the test set. The validation scores sourced from the leaderboard and test scores provided by the organizers. Our EASR-DCN shows superior performance in most metrics. Compared to TM-Large \cite{chen2022transmorph}, our EASR-DCN achieves a 4.6\% higher DSC, 0.422 lower HdDist95, and 0.084 lower SDlogJ in the validation set, reflecting improved registration accuracy with fewer DVF folds. These advantages are further amplified in the test set, where our EASR-DCN achieves a 6.5\% improvement in DSC, along with reductions of 0.358 in HdDist95 and 0.077 in SDlogJ.

The experimental results demonstrate the robust performance of our EASR-DCN in the MICCAI Learn2Reg 2021 Challenge (Task 3), thus confirming its competitiveness under standardized evaluation conditions.
}
%
\begin{table}
\centering
\scriptsize
\caption{Distribution of voxel intensity for different $k$-values.}
\label{DSC-ASD}
\begin{tabular}{|c|c|c|c|c|}
\hline
\rowcolor{gray! 10} {Dataset }&$k$-value& Anatomical Structure
 & Intensity Distribution  \\ \hline
 \multirow{3}*{\makecell[c] {OASIS Brain MR}} & \multirow{3}*{\makecell[c] { $k$=3}} & GM+Cortex & $(0, 0.18)$  \\ 
 \cline{3-4}
& & WM & $(0.18, 0.57)$  \\ 
\cline{3-4}
 & & CSF+IR   & $(0.57, 0.89)$   \\ \hline
%
%
\multirow{3}*{\makecell[c] {Hippocampus MR}} & \multirow{3}*{\makecell[c] { $k$=3}}  & IR1 & $(0, 0.15)$  \\ 
 \cline{3-4}

& &  Anterior + Posterior  & $(0.15, 0.69)$  \\ 
\cline{3-4}
 & & IR2   & $(0.69, 0.77)$   \\ \hline
%
%
\multirow{3}*{\makecell[c] {Cardiac CT}} & \multirow{3}*{\makecell[c] { $k$=3}} &  RA+LA+PA+Myo  & $(0, 0.39)$  \\ 
\cline{3-4}
 & & AO+LV & $(0.39, 0.67)$  \\ 
 \cline{3-4}
 & & RV+IR   & $(0.67, 0.81)$   \\ \hline
%
%
 \multirow{4}*{\makecell[c]{Cardiac MR}} & \multirow{4}*{\makecell[c]{$k$=4}} & RA+RV  & $(0, 0.19)$ \\ 
 \cline{3-4}
 & & PA & $(0.17,0.36)$  \\ 
 \cline{3-4}
 & & AO+LA+LV    & $(0.36,0.58)$    \\ 
 \cline{3-4}
 & & Myo+IR  & $(0.58,0.73)$  \\ 
 \hline
\end{tabular}
\label{voxeldistribution}
\end{table}
%
%
\begin{table*}
\centering
\setlength\tabcolsep{6 pt}
\caption{Four Types of Dynamic ROI Combination Scenarios and Their Typical Clinical Applications.
}
\begin{tabular}{|c|c|c|}
\hline 
\rowcolor{gray! 10} ROI type & Basis of the assumptions & Clinical application\\
  \hline 
Single structure &	\makecell[c]{Morphological, intensity homogeneity (\textit{e.g.}, white matter)}	& \makecell[c]{Microstructural analysis of white matter lesions }\\
 \hline 
 Adjacent-structure merge &Spatial continuity (\textit{e.g.}, RA/RV Shared Wall)	&Calculation of overall cardiac ejection fraction\\
\hline 
 Cross-Region functional merge&	Functional synergy (\textit{e.g.}, Ao/LV hemodynamics)	&Quantitative analysis of aortic valve regurgitation\\
\hline 
Anatomical structure + IR&Exclusion of non-functional signals (\textit{e.g.}, CSF))&Improved brain volumetry \\
\hline 
\end{tabular}
\label{assumption}
\end{table*}
%
\subsection{Analysis of $k$-value}
In EASR-DCN, the parameter $k$ (number of Gaussian components) determines effective ROI quantity and significantly impacts registration performance. 
Through systematic evaluation across three datasets (Table \ref{kanalysis}), we incrementally varied $k$ from 1 to 5, observing that performance initially improved and then declined as $k$ exceeded the optimal value corresponding to the actual ROI counts. The results revealed that $k$=3 is optimal for OASIS Brain MRI and Hippocampus MRI. Cardiac MRI achieved peak performance at $k$=4, with higher values ($k$ = 5) degrading the results in all cases. These findings informed our final parameter selection: $k$ = 3 for the brain / hippocampus and $k$ = 4 for the cardiac.

{\color{black}The optimal $k$-value in our EASR-DCN varies between datasets due to anatomical heterogeneity, which requires a modality-specific determination to improve performance. As illustrated in Table \ref{voxeldistribution}, brain MRI naturally clusters into 3 intensity-based ROIs (WM, GM+Cortex, IR+CSF), cardiac MRI requires 4 ROIs (RA+RV, AO+LA+LV, PA, Myo) to distinguish blood pools and myocardial tissue. Similarly, cardiac CT achieves optimal separation with 3 ROIs (AO+LV, PA+Myo+RA+LA, RV+IR), and hippocampus MRI divides into 3 regions (Anterior+Posterior, IR1, IR2). As illustrated in Table \ref{assumption}, these organ-specific partitions reflect clinically meaningful tissue contrasts and structural complexity. To streamline $k$ selection for new datasets, we propose a computationally efficient three-phase adaptive strategy that combines anatomical priors with data-driven validation, avoiding exhaustive brute-force searches while ensuring registration accuracy.} 
\begin{itemize}
\item {\color{black}Phase 1: Coarse Anatomical Prior Estimation}

{\color{black}We set biologically plausible $k$-ranges based on each modality's anatomical structures: Cardiac MR/CT: $k \in [1,7]$ (RA,RV,AO,LA,LV,PA,Myo); Brain MR: $k \in [1,4]$ (Cortex,WM,GM,CSF); Hippocampal MR: $k \in [1,3]$ (anterior/posterior,non-functional regions).}
\item {\color{black}Phase 2: Intensity-based Pre-screening}

{\color{black} Within the established anatomical prior ranges, we perform automated histogram analysis of the voxel intensity distributions. Key characteristics of the intensity profiles, such as modality-specific bimodality (\textit{e.g.}, CSF vs WM vs Cortex+GM in T1 brain MRI suggesting $k$=3), multi-compartment contrast (\textit{e.g.}, blood pool vs myocardium vs fat in cardiac MRI suggesting $k$ values between 3 and 4), or subfield homogeneity (\textit{e.g.}, hippocampal subfield intensity gradients suggesting $k$ values of 2 or 3), guide the selection of a pre-screened candidate $k$-value.}
{\color{black}\item {Phase 3: Fine-grained Validation with Local Search}}

{\color{black}This phase conducts a local search within a range of $k$ $\pm$ 1 around the pre-screened candidate. We use a structure-specific DSC evaluated on a small holdout set (5\%) as the validation. Specific termination criteria are used to avoid unnecessary computations: the process terminates when increasing $k$ by 1 ($k$+1) reduces the DSC by more than 10\% in critical structures, or when decreasing $k$ by 1 ($k$-1) improves the DSC by less than 5\% in critical structures. The search range is expanded only if performance remains stable within the initial local search.}
\end{itemize}
{\color{black}In the testing stage, we use the optimal $k$-value to guide GMM for segmentation to support downstream registration tasks.}
%
\begin{table}
\centering
\scriptsize
\setlength\tabcolsep{1 pt}
\caption{Evaluation of GMM Segmentation for our EASR module on the three  public MRI datasets. The {\color{red} {red numbers}} denote the effective segmentation results.
}
\begin{tabular}{|c|c|c|c|c|c|c|ccccccccc}
\hline
\rowcolor{gray! 10}Dataset &\multicolumn{2}{c|}{OASIS Brain MRI}&\multicolumn{2}{c|}{Hippocampus MRI}&\multicolumn{2}{c|}{Cardiac MRI} 	\\
\hline
 $k$-value &~~ DSC (\%) & ~~HdDist95 & DSC (\%) ~~ &HdDist95~~  &DSC (\%)  ~~&~~HdDist95 \\
\hline 
1 & $98.79$ & $0.106$ & $99.03$ & $0.097$ & $99.42$ & $0.085$  \\
%
2  & $95.79$ & $0.931$ & $92.66$ & $0.995$ & $93.30$ & $2.167$   \\
3 & \color{red} \color{red}$92.36$ & \color{red} $1.168$ & \color{red} $87.30$ & \color{red} $1.674$ & $90.64$ & $2.910$   \\
4 & $60.08$ & $8.551$ & $58.69$ & $8.244$ & \color{red} $88.20$ & \color{red} $3.209$  \\
 5 & $48.30$ & $12.260$ & $46.70$ & $11.929$ & $52.07$ & $9.957$  \\
\hline
\end{tabular}
\label{SRanalysis}
\end{table}
%
%
\begin{table}
\centering
\scriptsize
\setlength\tabcolsep{9 pt}
\caption{Segmentation results of our EASR module for OASIS Brain MRI, Hippocampus MRI, and Cardiac MRI datasets are reported under optimal $k$-values.}
\label{DSC-ASD}
\begin{tabular}{|c|c|c|c|}
\hline
\rowcolor{gray! 10} {Dataset ($k$-value)}&{ROI}&{DSC (\%)} &{HdDist95}  \\ \hline
\multirow{3}*{\makecell[c] {OASIS Brain MRI\\($k$=3)}} & GM+Cortex  & $93.80 ~(0.92)$ & $0.171$ \\ 
\cline{2-4}
 & WM & $92.44~(1.07)$ & $0.169$ \\ 
 \cline{2-4}
 & CSF   & $90.84 ~(1.64)$  & $0.197$  \\ \hline
%
%
 \multirow{2}*{\makecell[c]{Hippocampus MRI \\($k$=3)}} & Anterior  & $88.92 ~ (1.77)$ & $1.583$ \\ 
 \cline{2-4}
 &Posterior & $85.68~(2.09)$ & $1.755$ \\ \hline
%
%
 \multirow{4}*{\makecell[c]{Cardiac MRI \\($k$=4)}} & RA+RV  & $90.08~(1.66)$ & $2.968$ \\ 
 \cline{2-4}
 & AO+LA+LV & $89.76 ~(1.83)$ & $2.977$ \\ 
 \cline{2-4}
 & PA   & $86.95 ~(1.94)$  & $3.205$  \\ 
 \cline{2-4}
 & Myo  & $86.01 ~(1.97)$ & $3.371$ \\ 
 \hline
\end{tabular}
\label{SR}
\end{table}
%
%
\begin{figure}
\center
\includegraphics[width=0.47\textwidth]{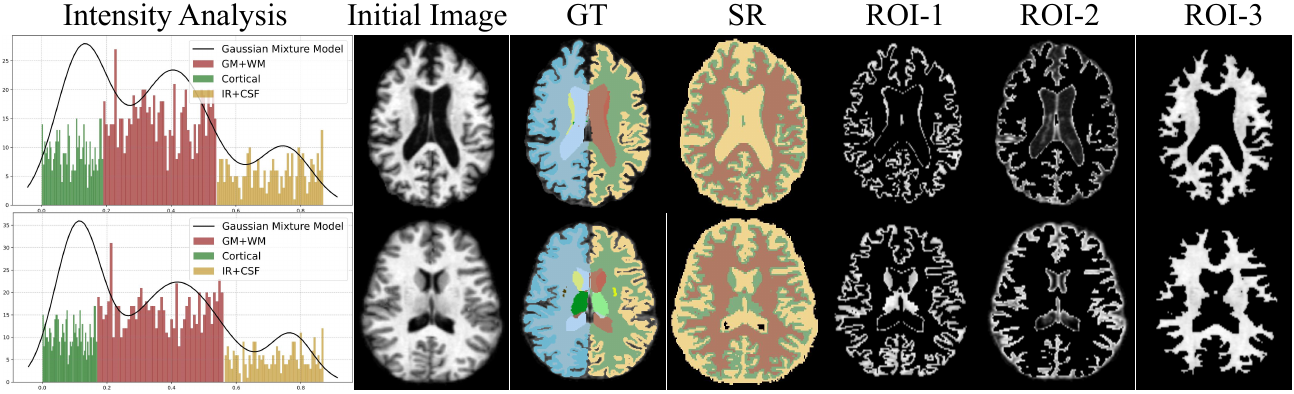}
\includegraphics[width=0.47\textwidth]{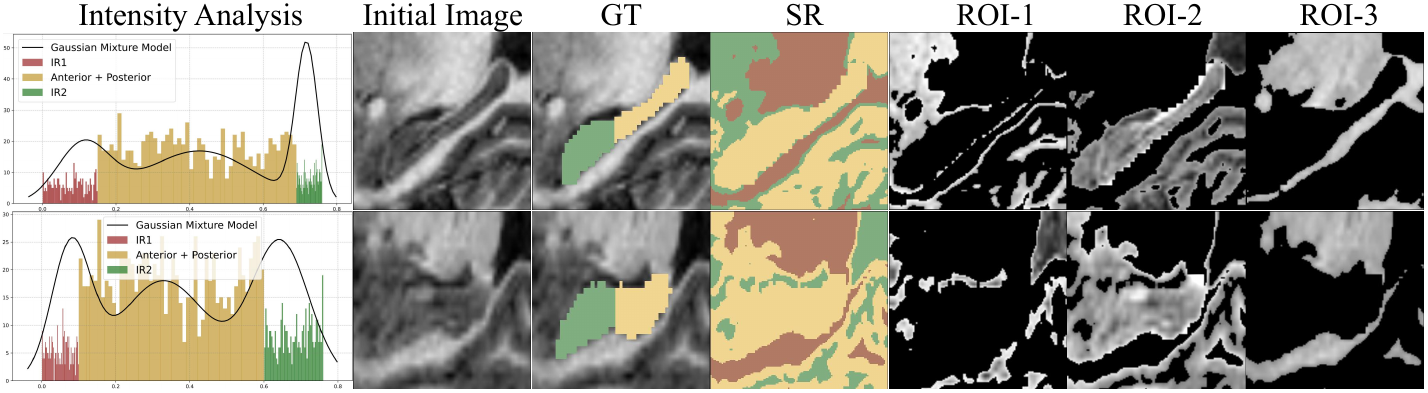}
\includegraphics[width=0.47\textwidth]{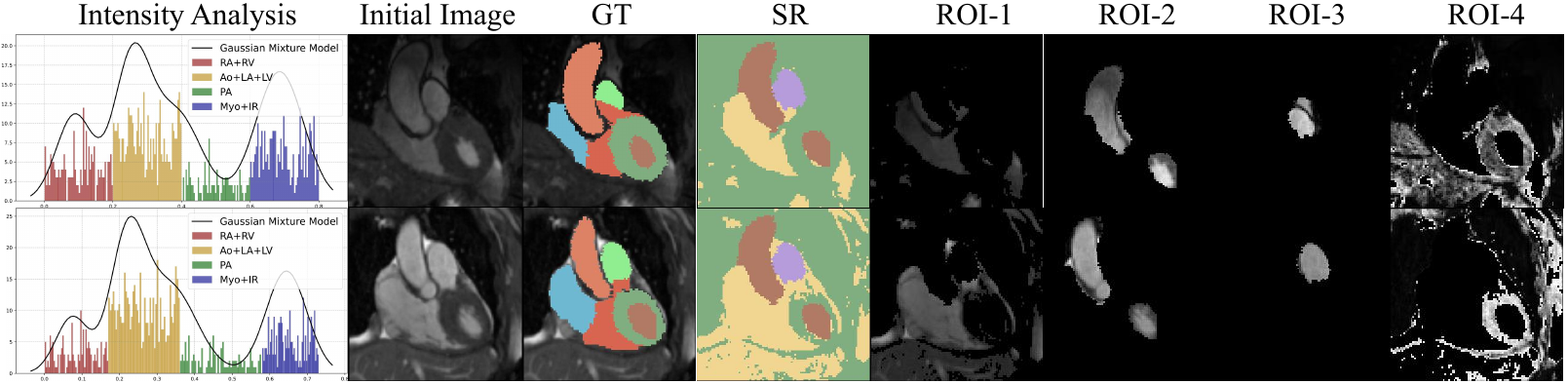}
\caption{Segmentation results of effective ROIs from three MRI datasets: OASIS Brain MR, Hippocampus MR, and Cardiac MR. The GT denotes Ground Truth, and SR denotes the Segmentation Result.}
\label{demoGMMSeg}
\end{figure}
%
\begin{table*}
\centering
\scriptsize
\caption{Ablation analysis on three MRI datasets, with the DCN operating under optimal $k$-value and evaluated using DSC (\%), $|{J}_{\phi}|{\leq 0}$ (\%), and HdDist95 as metrics. The bold indicate the highest scores. Folds are presented in e-notation (e.g., \(1\textit{e}{-2} = 0.01\)).
}
\begin{tabular}{|c|c|c|c|c|c|c|c|c|c|c|c|}
\hline
\rowcolor{gray! 5}{-}&\multicolumn{2}{c|}{Optional}&\multicolumn{3}{c|}{OASIS Brain MRI}&\multicolumn{3}{c|}{Cardiac MRI}&\multicolumn{3}{c|}{Hippocampus MRI}\\
\hline
\rowcolor{gray! 5}{Method} &{ROI-based} &{DCN}& DSC (\%) & $|{J}_{\phi}|{\leq 0}$ (\%) & HdDist95  & DSC (\%) & $|{J}_{\phi}|{\leq 0}$ (\%) & HdDist95& DSC (\%) & $|{J}_{\phi}|{\leq 0}$ (\%) & HdDist95 \\
\hline
ScU \cite{balakrishnan2019voxelmorph} & \XSolidBrush &\XSolidBrush&$78.89$ &$5.57e{-2}$& $2.107$ & $75.69$  & $6.28e{-2} $  & $5.306$ & $74.29$  & $5.92e{-2} $ & $ 2.753 $\\
SAMReg \cite{kirillov2023segment} &\CheckmarkBold &\XSolidBrush& $88.95$ & $1.56e{-2}$ & $1.935$ & $88.37$  & $1.06e{-2} $ & $ 3.597$ & $79.96$  & $1.67e{-2} $ & $2.485$\\
EASR-DCN &\CheckmarkBold & \CheckmarkBold & $\textbf{89.20}$ &$\textbf{1.50e{-2}}$& $\textbf{1.924}$ & $\textbf{88.70}$  & $\textbf{1.03e{-2}} $ & $ \textbf{3.588}$ & $\textbf{80.04}$  & $\textbf{1.62e{-2}} $& $ \textbf{2.397} $\\
\hline 
\end{tabular}
\label{ablation}
\end{table*}
%
\subsection{Evaluation of GMM Segmentation}
In our EASR-DCN model, GMM segmentation guides subsequent registration. We measured its impact on accuracy and anatomical preservation using DSC and HdDist95.

Table \ref{SRanalysis} shows the average value of DSC and HdDist95 in three datasets. When \( k = 1 \), where the entire image is treated as a single ROI, registration achieves perfect alignment with ground truth, resulting in near-zero error. As \( k \) increases, the accuracy of the GMM segmentation decreases, leading to higher registration errors. However, within an optimal \( k \) range, GMM remains effective: for OASIS Brain and Hippocampus MRI, performance is strong with \( k \) values from 1 to 3, but decreases at \( k = 4 \) and 5. For cardiac MRI, accuracy persists for \( k \) values up to 4 but deteriorates at \( k = 5 \). These trends correlate with the registration performance in Table \ref{kanalysis}.  

Table \ref{SR} presents the results of segmentation using optimal $k$-values: three ROIs for OASIS brain MRI ( GM+Cortex, WM, IR+CSF), three for Hippocampus MRI (anterior+posterior, IR1, IR2), and four for cardiac MRI (RA+RV, AO+LA+LV, PA, Myo+IR). GMM achieves high segmentation accuracy across all datasets.

Fig. \ref{demoGMMSeg} shows the segmentation performance of GMM. In OASIS brain MRI, it precisely separates CSF+IR, GM+Cortex, and WM. For Hippocampus MRI, it clearly differentiates anterior + posterior, IR1 and IR2 regions. Cardiac MRI segmentation effectively identifies RA+RV, AO+LA+LV, PA, and Myo+IR, where multiple anatomical structures are combined within single ROIs.

In summary, GMM segmentation effectively preserves anatomical structures and provides accurate segmentation results, thus offering valuable guidance for downstream registration tasks and enhancing overall registration performance.
\begin{figure}
\centering
\includegraphics[width=0.40\textwidth]{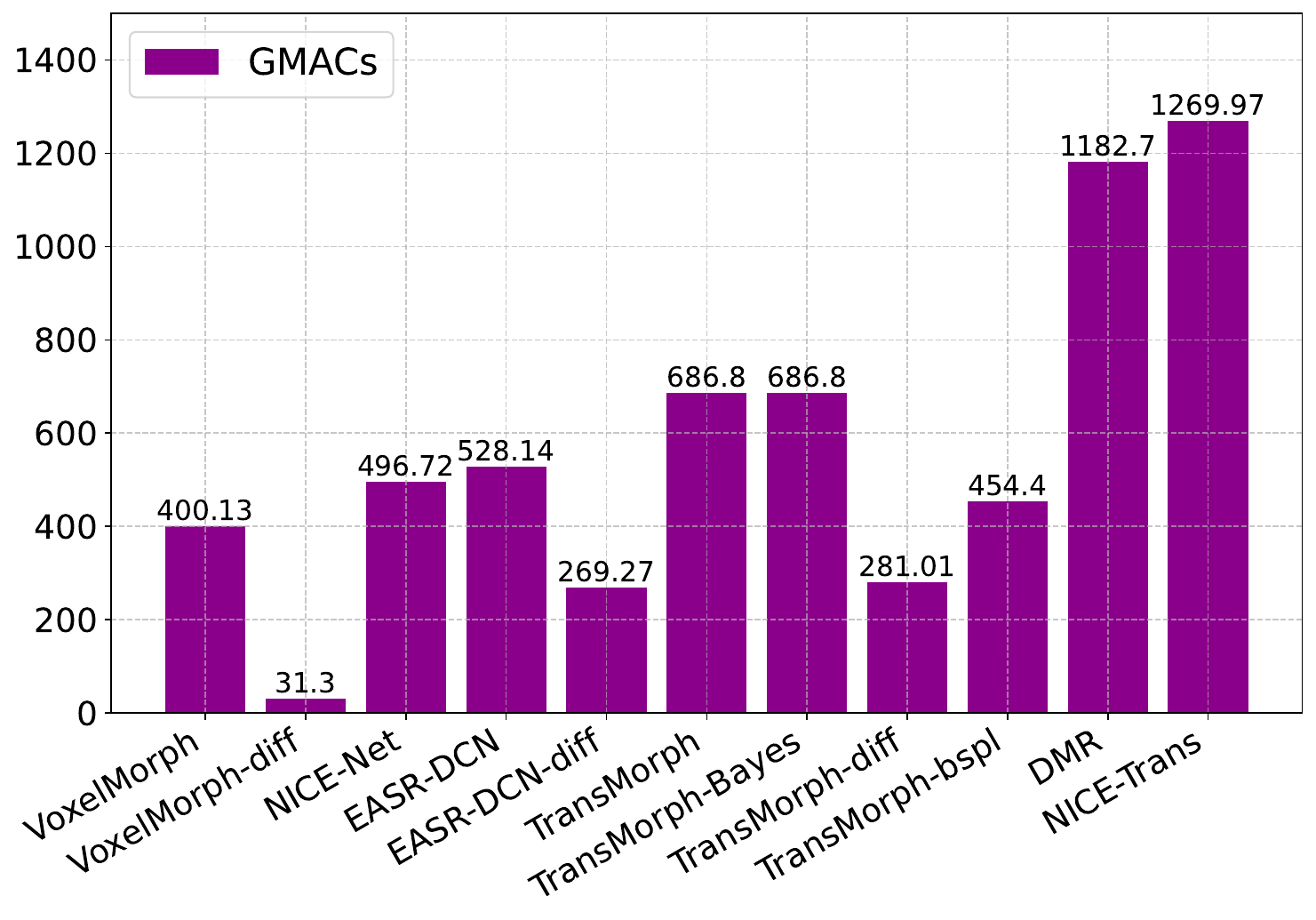}
\caption{Comparison of model computational complexity in Giga Multiply-Accumulate Operations (GMACs). Higher values indicate greater complexity. Measured with $160 \times 192 \times 224$ voxel input images.}
\label{complexity}
\end{figure}
\begin{figure}
\centering
\includegraphics[width=0.40\textwidth]{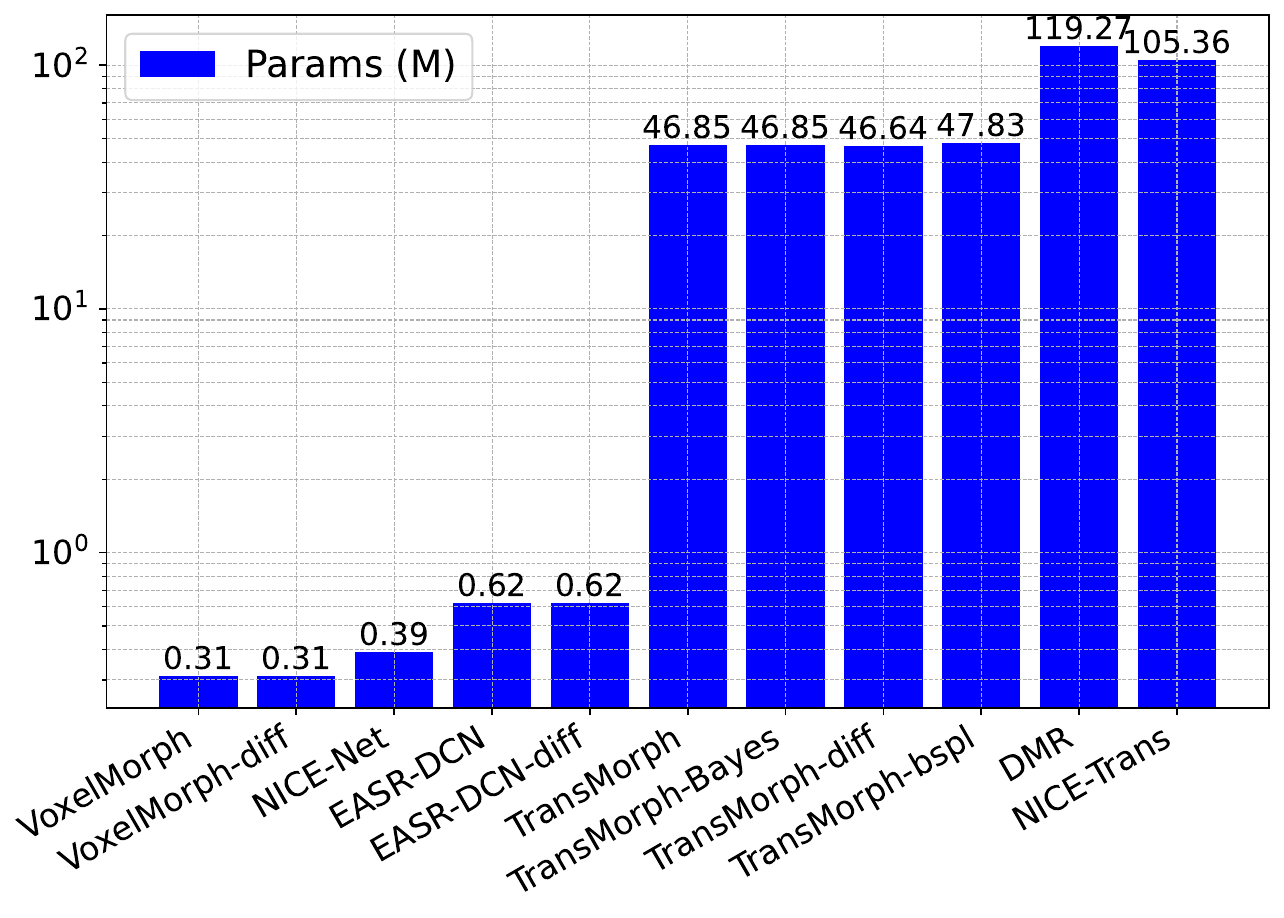}
\caption{Number of trainable parameters (in millions) for each learning-based model.}
\label{paramas}
\end{figure}
%
%
\begin{table}
\centering
\setlength\tabcolsep{4 pt}
\caption{Average training and inference time for compared methods used in this work. SyN \cite{avants2008symmetric} was applied using CPUs, and the learning-based methods were implemented on GPU. Inference time was averaged based on 50 repeated runs.
}
\begin{tabular}{|c|c|c|}
\hline 
\rowcolor{gray! 10} Method & Training (min/epoch) & Inference (sec/image)\\
  \hline 
 SyN \cite{avants2008symmetric} & - & 192.14\\
 \hline 
 VoxelMorph \cite{balakrishnan2019voxelmorph} & 9.40 & 0.43\\
\hline 
 VoxelMorph-diff \cite{balakrishnan2019voxelmorph} & 4.20 & 0.05\\
\hline 
 TransMorph \cite{chen2022transmorph} & 14.40 & 0.33\\
\hline 
 TransMorph-diff \cite{chen2022transmorph} & 7.35 & 0.10\\
\hline 
 TransMorph-Bayes \cite{chen2022transmorph} & 22.60 & 7.74\\
\hline 
 TransMorph-bspl \cite{chen2022transmorph} & 10.50 & 1.74\\
\hline 
 NICE-Net \cite{meng2022non} & 9.93 & 0.95\\
\hline 
 DMR \cite{chen2022deformer} & 18.02 & 0.63\\
\hline 
 NICE-Trans \cite{meng2023non} & 19.25 & 0.76\\
\hline 
 EASR-DCN (ours) & 13.27 & 0.84 \\
\hline 
 EASR-DCN-diff (ours) & 9.95 & 0.38\\
\hline 
\end{tabular}
\label{runtime}
\end{table}
%
\subsection{Ablation Study}
Table \ref{ablation} compares the registration performance between our EASR-DCN, SAMReg \cite{huang2024one} and ScU \cite{balakrishnan2019voxelmorph}. Both EASR-DCN and SAMReg outperform ScU, and our method shows superior performance among ROI-based approaches.

Compared to SAMReg \cite{huang2024one}, our EASR-DCN achieves DSC improvements of 0.25\% (OASIS Brain MRI), 0.33\% (Cardiac MRI), and 0.08\% (Hippocampus MRI), while reducing HdDist95 by 0.011, 0.009, and 0.088 respectively. It also decreases $|{J}_\phi| \leq 0$ instances by 0.06\%, 0.03\%, and 0.05\%.

In general, our DCN is an effective registration network that improves registration accuracy and reduces DVF folds.
\subsection{{Computational complexity}} 
{\color{black}To provide a comprehensive evaluation, we conducted a detailed comparison between our EASR-DCN and five SOTA registration methods, focusing on memory usage and computational complexity. The experimental setup involved implementing our EASR-DCN in PyTorch on an NVIDIA RTX3090 GPU and training models for 500 epochs using the Adam optimizer.
We analyze the computational complexity in terms of Giga Multiply-Accumulate Operations (GMACs) and the number of trainable parameters. The results were obtained using an input image with a resolution of 160 $\times$ 192 $\times$ 224 from OASIS Brain MR images. As shown in Fig. \ref{complexity}, our EASR-DCN has moderate computational complexity (528.14 GMACs) compared to Transformer models like TransMorph \cite{chen2022transmorph}, DMR \cite{chen2022deformer}, and NICE-Trans \cite{meng2023non}, which have substantially higher GMAC counts. The diffeomorphic variant, EASR-DCN-diff, is even more computationally efficient (269.27 GMACs). Regarding memory usage during training, our EASR-DCN occupied approximately 13 GB of GPU memory with a batch size of 1 and an input image size of 160 $\times$ 192 $\times$ 224, while our EASR-DCN-diff required about 10 GiB, sizes readily accommodated by most modern GPUs.

Fig. \ref{paramas} illustrates the number of trainable parameters. All ConvNet-based models have fewer than 1M parameters, yet their GMACs are comparable to our EASR-DCN, despite significantly inferior registration performance. Transformer-based models have a substantially larger scale, with parameter counts exceeding 45M, and some (DMR \cite{chen2022deformer}, NICE-Trans \cite{meng2023non}) surpassing 100M. Notably, our EASR-DCN outperformed these larger models across evaluated tasks. This demonstrates that the superior performance of our EASR-DCN stems not merely from model size but rather from its effective network design.
}  
\subsection{{Run-time Analysis}} 
{\color{black}Table \ref{runtime} provides a comparative analysis of runtime in all methods, benchmarking both the training duration per epoch (min / epoch) and the inference latency per image (sec/image). Notably, SyN \cite{avants2008symmetric} was implemented on CPU architecture, whereas all deep learning-based frameworks utilized GPU acceleration. Performance metrics were standardized using input dimensions of 160 $\times$ 192 $\times$ 224 voxels, matching the spatial resolution characteristics of the OASIS brain datasets. }

{\color{black}The per-epoch training duration was determined using a dataset comprising 768 image pairs. Among all evaluated methods, the two Transformer-based architectures, TransMorph-Bayes \cite{chen2022transmorph} and NICE-Trans \cite{meng2023non}, demonstrated the greatest computational demands. Specifically, TransMorph-Bayes \cite{chen2022transmorph} required approximately 7.85 days to complete 500 training epochs, calculated as (22.60 min/epoch × 500 epochs)/(60 min/hour × 24 hours/day). NICE-Trans \cite{meng2023non} exhibited the second highest temporal cost, needing 6.68 days for equivalent training, derived from (19.25 min/epoch × 500 epochs)/(60 × 24). TransMorph-Bayes \cite{chen2022transmorph} exhibited significant computational overhead due to its requirement for repeated sampling of training images during both individual predictions and uncertainty quantification processes. Furthermore, the training time of NICE-Trans \cite{meng2023non} slowed because of the computationally intensive nature of its extensive convolutional operations. The proposed EASR-DCN has a moderate training speed, required approximately 4.61 days to complete 500 training epochs, calculated as (13.27 min/epoch × 500 epochs)/(60 min/hour × 24 hours/day).}

{\color{black}In terms of inference time, learning-based models demonstrably surpass traditional registration methods by orders of magnitude. Among learning-based registration approaches, only TransMorph-Bayes \cite{chen2022transmorph} and TransMorph-bspl \cite{chen2022transmorph} exceed the 1-second threshold, while all other methods complete within sub-second intervals. Notably, our proposed EASR-DCN and EASR-DCN-diff achieve inference times of 0.84 seconds and 0.38 seconds respectively, showcasing the model's capacity for real-time registration performance.}
\subsection{{Noise Sensitivity Testing}} 
\begin{figure}
\centering
\includegraphics[width=0.40\textwidth]{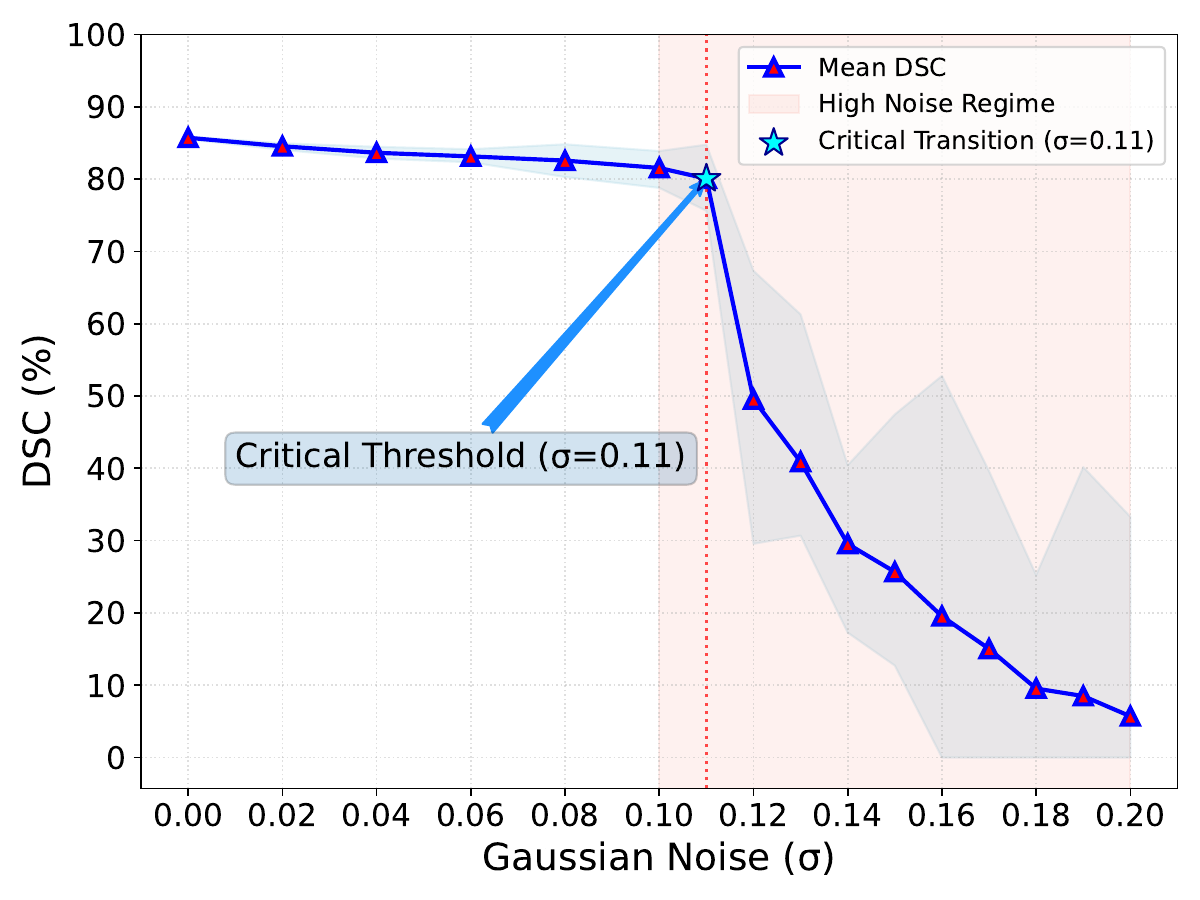}
\caption{Noise sensitivity test. We simulate Gaussian noise to conduct noise testing on the input images, with the noise variance $\sigma$ set to [0, 0.2]. Registration performance is measured using DSC. The pink region indicates the high-noise area, while the blue shaded region represents the variance in registration accuracy.}
\label{noise}
\end{figure}

{\color{black}Concerning sensitivity to noise, we conducted comprehensive noise sensitivity tests by adding Gaussian noise with varying intensities ($\sigma$ ranging from 0 to 0.2) to the Cardiac CT test set, covering both clinical and extreme cases. As illustrated in Fig. \ref{noise}, our EASR-DCN maintains stable registration accuracy (DSC $\textgreater$ 0.80) within clinically relevant noise levels ($\sigma$  $\textless$ 0.11). While performance degrades at higher noise levels, which is consistent with the fundamental limits of intensity-based registration, the current implementation demonstrates satisfactory robustness to moderate noise. We acknowledge that further enhancing noise robustness through techniques like denoising preprocessing or feature-based registration would be valuable for extremely noisy scenarios, and this is an important direction for our future research.
}
\begin{table}
\centering
\scriptsize
\setlength\tabcolsep{2pt}
\caption{Hyper-parameter study of our EASR-DCN on three MRI datasets. DSC and $|{J}_{\phi}|{\leq 0}$ were used as evaluation metrics. Folds are presented in e-notation (e.g., \(1\textit{e}{-2} = 0.01\)).
}
\begin{tabular}{|c|c|c|c|c|c|c|c|}
\hline
\rowcolor{gray! 10} Dataset& Metric & $\alpha$ = 0 & $\alpha$ = 1 & $\alpha$ = 2 & $\alpha$ = 3 & $\alpha$ = 4 & $\alpha$ = 5 \\
\hline
 \multirow{2}*{\makecell[c] {OASIS Brain  \\$k$ = 3}} & DSC (\%)&  70.98 &  91.14 & 90.03  & 89.20 & 87.64 & 86.92\\
 & $|{J}_{\phi}|{\leq 0}$ (\%) & 1.39\textit{e}{-1} & 9.39\textit{e}{-2} & 6.67\textit{e}{-2} & 1.50\textit{e}{-2}  &  6.19\textit{e}{-3} & 2.07\textit{e}{-3}   \\
\hline 
 \multirow{2}*{\makecell[c] {Hippocampus \\$k$ = 3}} 
 & DSC ($\%$) &  71.35 &  80.04 & 79.26  & 78.05  & 78.03 & 78.00 \\
 &$|{J}_{\phi}|{\leq 0}$ (\%) & 2.07\textit{e}{-1} & 1.62\textit{e}{-2} & 1.55\textit{e}{-2} & 1.02\textit{e}{-2}  & 4.38\textit{e}{-3}  & 3.17\textit{e}{-3}\\
\hline 
 \multirow{2}*{\makecell[c] {Cardiac\\$k$ = 4}}& DSC (\%) & 61.78 & 90.78& 89.03 & 88.70  & 87.29& 87.22 \\
 & $|{J}_{\phi}|{\leq 0}$ (\%) & 1.85\textit{e}{-1} & 9.96\textit{e}{-2} & 7.87\textit{e}{-2} & 1.06\textit{e}{-2}  & 8.95\textit{e}{-3} & 3.89\textit{e}{-3}\\
\hline
\end{tabular}
\label{hyperparameter}
\end{table}
\subsection{{Hyper-parameter Study}} 
The hyper-parameter $\alpha$ serves as a regularization parameter in our EASR-DCN model. Same as VM \cite{balakrishnan2019voxelmorph}, we varied $\alpha$ within the range $[0, 5]$ to assess its impact on registration performance, measured by average DSC and the percentage of voxels with a non-positive Jacobian determinant $|{J}_\phi| \leq 0$.

As shown in Table \ref{hyperparameter}, average DSC values slightly decrease as $\alpha$ increases while the percentage of voxels with $|{J}_\phi| \leq 0$ gradually approaches zero, indicating that EASR-DCN is relatively robust to $\alpha$ settings. For optimal DSC values, a smaller value of $\alpha$ within the range [1,5] is preferable, while for minimizing $|{J}_\phi| \leq 0$, a larger $\alpha$ is favored.

Empirically, we set $\alpha$ to 3 for the OASIS Brain MRI and Cardiac MRI datasets and 1 for the Hippocampus MRI dataset to balance registration accuracy with fold elimination.
\section{Discussion and Limitation}
\label{Discussion and Limitation}
\begin{figure}[h]
\centering
\includegraphics[width=0.47\textwidth]{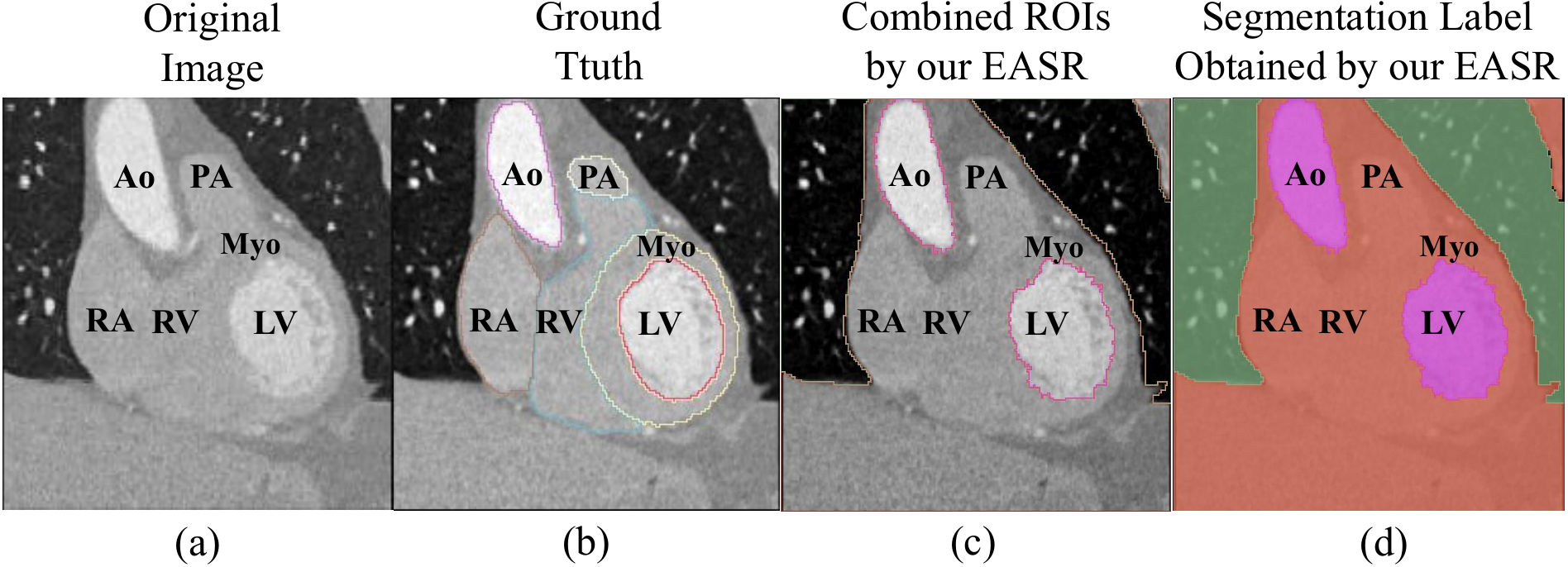}
\caption{Cardiac CT images. (a) Original CT image; (b) Ground truth of anatomical structures in the Cardiac CT image; (c) Segmentation results obtained using our EASR; (d) Label-map obtained by our EASR.}
\label{cardiacCT}
\end{figure}

\begin{figure}[H]
\centering
\includegraphics[width=0.47\textwidth]{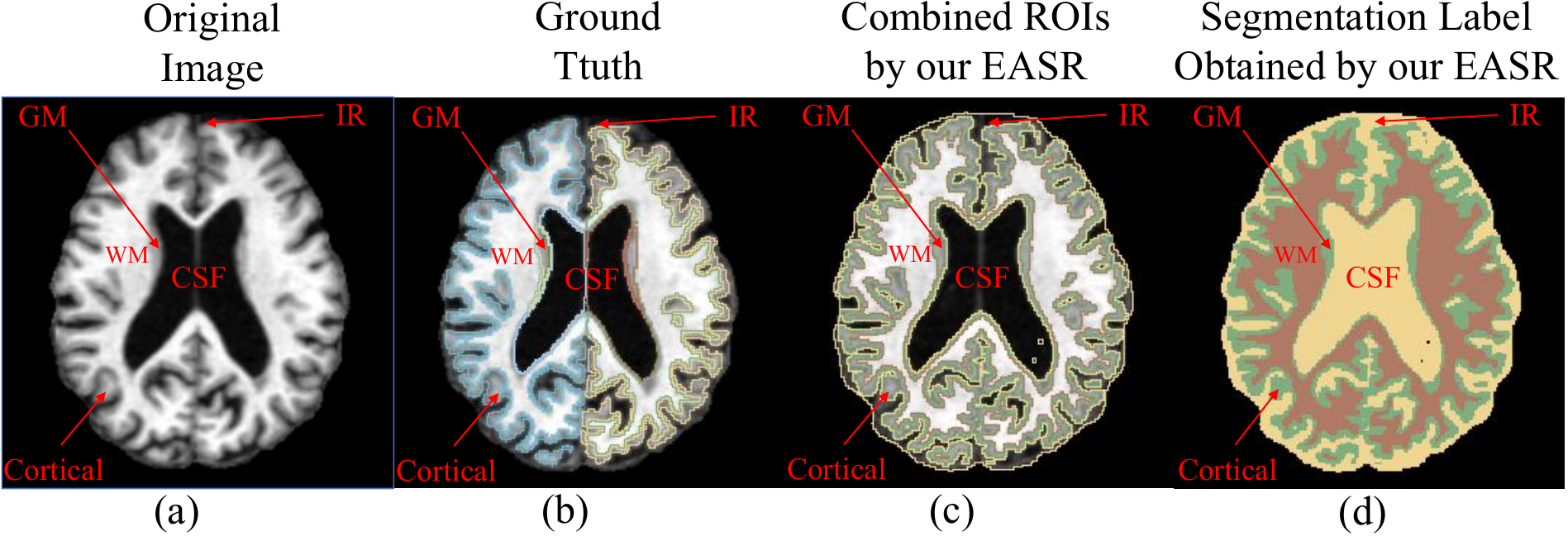}
\caption{Brain MR Images. (a) Original Brain MR image; (b) Ground truth of anatomical structures in the Brain MR image; (c) Segmentation results obtained using our EASR; (d) Label-map obtained by our EASR.}
\label{brainMR}
\end{figure}

{\color{black}In this paper, we recognize that relying solely on intensity can be challenging due to factors such as overlapping Hounsfield unit values in CT images and intensity inhomogeneities in MR images, as illustrated in Fig. \ref{cardiacCT}(a)-(b) and Fig. \ref{brainMR}(a)-(b). These limitations can indeed make it difficult to distinguish certain anatomical structures or separate irrelevant regions from relevant ones based on intensity alone.} 

{\color{black}Although we face these challenges, our approach operates under the assumption that, by leveraging both image intensity and anatomical consistency, it is possible to delineate effective ROIs. This allows us to move beyond strict classical anatomical boundaries and dynamically reorganize ROIs based on the requirements of the downstream registration task. This assumption is supported by several factors: the distinct characteristics of human organs in terms of morphology, position, and function; the consistency of intensity distributions within certain tissue types (\textit{e.g.}, uniform grayscale in WM); spatial continuity, such as shared walls between adjacent cardiac chambers; and functional synergy observed between certain regions ( \textit{e.g.}, hemodynamic coupling of the aorta and LV).} 

{\color{black}\textbf{Limitation:} Despite its advantages, we acknowledge certain limitations. The proposed EASR-DCN is mainly suited for unimodal MR or CT registration, where distinct intensity differences facilitate ROI segmentation. It faces challenges in multimodal registration scenarios involving missing correspondences, as the EASR strategy may not reliably identify valid ROI pairs. Furthermore, under conditions of extreme noise, the number of reliable ROIs that can be extracted may decrease, potentially leading to degraded registration performance.}

{\color{black}\textbf{Future Work:} Our future work will focus on integrating shape priors and cross-modal feature descriptors to enable the framework's application to multimodal tasks.}
\section{Conclusion}
\label{Conclusion}
In this paper, we present EASR-DCN, an innovative unsupervised learning registration method tailored for DMIR applications. EASR-DCN efficiently captures ROIs within images, thereby enhancing the performance of subsequent registration tasks. Additionally, we introduce a novel DCN architecture that facilitates the unsupervised, independent alignment of ROIs between image pairs. Extensive experiments were performed on three MRI datasets and one CT dataset, encompassing various registration scenarios. The results demonstrate that our EASR-DCN eliminates the need for manual annotation. It significantly improves registration accuracy while reducing deformations in the DVFs. The EASR-DCN framework is adaptable to various applications and has shown substantial potential for future advancements in DMIR.


%

\section*{References}
\bibliographystyle{ieeetr}
\bibliography{Manuscript}
\end{document}